\documentclass[11pt,DIV=12,abstract=true,numbers=noenddot,titlepage=false,twocolumn=false,draft=false]{scrartcl}
\pdfoutput=1

\makeatletter
\DeclareOldFontCommand{\rm}{\normalfont\rmfamily}{\mathrm}
\DeclareOldFontCommand{\sf}{\normalfont\sffamily}{\mathsf}
\DeclareOldFontCommand{\tt}{\normalfont\ttfamily}{\mathtt}
\DeclareOldFontCommand{\bf}{\normalfont\bfseries}{\mathbf}
\DeclareOldFontCommand{\it}{\normalfont\itshape}{\mathit}
\DeclareOldFontCommand{\sl}{\normalfont\slshape}{\@nomath\sl}

\usepackage[affil-it]{authblk}
\usepackage[numbers,sort&compress]{natbib}

\usepackage[dvips]{graphicx}

\usepackage{siunitx}
\usepackage{array,amsmath,amsthm,feynmf,mathtools}
\usepackage{multirow}
\usepackage{mathtools}
\usepackage{epsfig}
\usepackage{graphicx}
\usepackage{xcolor}
\usepackage{slashed}
\usepackage{amssymb}
\usepackage{bm}
\usepackage[export]{adjustbox}
\usepackage{enumitem}
\usepackage[colorlinks=true,linkcolor=blue,citecolor=blue]{hyperref}

\allowdisplaybreaks

\begin{document}

\renewcommand\Authands{, }

\title{Hadronic CP Violation in the 2HDM}

\date{\today}
\author[a]{Wolfgang~Altmannshofer%
        \thanks{\texttt{waltmann@ucsc.edu}}}
\author[b]{Joachim~Brod%
        \thanks{\texttt{joachim.brod@uc.edu}}}
\author[c]{Patipan~Uttayarat%
        \thanks{\texttt{patipan@g.swu.ac.th}}}
\author[b]{Daniil~Volkov%
        \thanks{\texttt{volkovdi@mail.uc.edu}}}
	\affil[a]{{\large Department of Physics and Santa Cruz Institute for Particle Physics\\ University of California, Santa Cruz, CA 95064, USA}}
	\affil[b]{{\large Department of Physics, University of Cincinnati, Cincinnati, OH 45221, USA}}
	\affil[c]{{\large Department of Physics, Srinakharinwirot University, 114 Sukhumvit 23rd Rd., Wattana, Bangkok 10110, Thailand}}

\maketitle

\begin{abstract}
  We present the first complete two-loop calculation of the electric
  and chromo-electric dipole moments of the light quarks and the
  gluon, as well as contributions to CP-violating lepton-quark
  interactions, in the unconstrained two-Higgs doublet model. We
  include the most general Yukawa interactions of the Higgs doublets
  with the Standard Model fermions up to quadratic order, and allow
  for generic phases in the Higgs potential. We pay particular
  attention to a consistent treatment of all fermionic contributions
  in the low-energy effective theory, including a consistent
  renormalization-group summation of all leading-logarithmic
  effects. This latter part of the work is independent of the specific
  UV model and can generally be applied to a large class of models
  that do not introduce new light degrees of freedom. A
  \texttt{python} implementation of our results is provided via a
  public git repository.
\end{abstract}

\section{Introduction}

Bounds on electric dipole moments (EDMs) provide stringent constraints
on CP-violating phases that are crucial for models of electroweak
baryogenesis~\cite{Kuzmin:1985mm}. In the two-Higgs doublet model
(2HDM)~\cite{Deshpande:1977rw}, the desired CP violation is supplied
by the complex phases associated with the Higgs sector that is
extended with respect to the SM by adding a second Higgs doublet. It
is mainly the phases in the Yukawa couplings to the third fermion
generation that play a role in models of electroweak
baryogenesis~\cite{Huber:2006ri, deVries:2017ncy, DeVries:2018aul,
  Fuyuto:2019svr, Ge:2020mcl, Fuchs:2020uoc, Bahl:2022yrs,
  Athron:2025iew}. The Higgs potential of the 2HDM allows for a
first-order phase transition for an appropriate choice of
parameters~\cite{Bernreuther:2002uj, Morrissey:2012db, Dorsch:2013wja,
  Basler:2017uxn, Andersen:2017ika, Bernon:2017jgv, Basler:2018cwe,
  Aoki:2021oez, Basler:2021kgq, Song:2022xts, Wagner:2023vqw,
  Anisha:2025zbc, RadchenkoSerdula:2025shh}.

The main constraints on the CP phases arise from the non-observation
of electric dipole moments (EDMs) of elementary systems such as the
electron, the neutron, or atomic and molecular
systems~\cite{Engel:2013lsa, Inoue:2014nva, Chupp:2017rkp,
  Pospelov:2025vzj}. If one allows for the presence of phases in all
the Yukawa couplings, however, it is easy to arrange for the bounds
from any individual EDM to disappear due to the cancellation of the
various contributions. It is therefore imperative to include as many
observables as possible when testing the CP phases of the 2HDM. In
Ref.~\cite{Altmannshofer:2024mbj} we have calculated the leading
contributions to the electron EDM, and lepton-number violating
radiative lepton decays; see also Ref.~\cite{Davila:2025goc} for
related work. Here, we focus on the EDMs of hadronic systems. In
particular, we calculate the leading contributions to the light-quark
(up, down, and strange) EDMs and chromo-EDMs, to the Weinberg operator
(the gluon chromo-EDM), and CP-violating electron-quark
interactions. These CP-violation quark and gluon interactions are not
directly observable, but they induce a large variety of observable
EDMs in elementary, atomic, and molecular systems; see
Ref.~\cite{Engel:2013lsa} for a review.

We match the 2HDM directly to an effective theory valid below the
electroweak scale, so that our results are valid for arbitrary Higgs
masses (as long as they are at or above the electroweak scale). We
then sum all leading large logarithms to all orders in the strong
coupling constant by calculating the QCD renormalization-group (RG)
evolution of the Wilson coefficients from the electroweak scale to the
hadronic scale of 2\,GeV. In fact, this RG evolution applies also to
extensions of the SM other than the 2HDM, as long as no new degrees of
freedom below the weak scale are introduced. Hence, this part of our
results should be useful beyond the scope of the current analysis.

This work is structured as follows. We briefly introduce the 2HDM and
establish our conventions in Sec.~\ref{sec:2hdm}. The effective
theories below the weak scale, as well as the anomalous dimensions
required for the RG evolution, are introduced in Sec.~\ref{sec:rg}.
Our main results are contained in Sec.~\ref{sec:match}, where we
provide explicit expressions for all matching conditions the weak
scale. Sec.~\ref{sec:light} contains a detailed discussion of the
summation of large logarithms that arise from small SM fermion
masses. Sec.~\ref{sec:conclusions} comprises a brief summary and a
link to a public git repository containing a \texttt{python}
implementation of our results. Finally, in App.~\ref{sec:fierz} we
collect some Fierz identities that were used in obtaining our results,
and in App.~\ref{sec:gauge} we give the gauge-dependent parts of the
loop functions for some subsets of Feynman diagrams (as discussed
below, all gauge dependence cancels when summing the contributions of
all diagrams).

\section{The 2HDM}\label{sec:2hdm}

To begin, we briefly review the general two-Higgs doublet model. This
serves mainly to establish our notation and conventions; a detailed
discussion can be found in Refs.~\cite{Davidson:2005cw, Haber:2006ue,
  Boto:2020wyf, Altmannshofer:2020shb}. We will work with the
unconstrained 2HDM in which the SM scalar sector is extended by an
additional scalar doublet, such that the scalar potential depends on
two $SU(2)_L$ doublets $H_1$ and $H_2$, both with hypercharge
$+1/2$. Here, we work in the the so-called Higgs basis, where only the
field $H_1$ receives a vacuum expectation value. In that basis, the
scalar potential is given by
\begin{equation}\label{eq:HBpot}
\begin{split}
\mathcal{V}(H_1,\,H_2)
& = Y_1 H_1^\dagger H_{1}
    + Y_2 H_2^\dagger H_{2}
    + \Big(Y_3  H_1^\dagger H_{2} +\text{c.c.}\Big) \\
& \quad + {\textstyle\frac{1}{2}}Z_1\big( H_1^\dagger H_{1}\big)^2
    + {\textstyle\frac{1}{2}}Z_2\big( H_2^\dagger H_{2}\big)^2
    + Z_3\big( H_1^\dagger H_{1}\big)\big( H_2^\dagger H_2\big)
    + Z_4\big( H_1^\dagger H_{2}\big)\big( H_2^\dagger H_1\big) \\
& \quad + \Big({\textstyle\frac{1}{2}}Z_5  \big( H_1^\dagger H_2\big)^2
    + \big(Z_6\, H_1^\dagger H_1 + Z_7\, H_2^\dagger H_2 \big)
      \big(H_1^\dag H_2\big) +\text{c.c.}\Big) \,,
\end{split}
\end{equation}
where the parameters $Y_1$, $Y_2$, $Z_1,\ldots,Z_4$ are real, while
$Y_3$ and $Z_5$, $Z_6$, $Z_7$ may be complex. In the Higgs basis, the
components of the scalar fields can be written as
\begin{equation}
H_1 =
\begin{pmatrix}
G^+\\
\frac{1}{\sqrt{2}}\big(v + \varphi_1^0 + i G^0\big)
\end{pmatrix}\,, \qquad 
H_2 =
\begin{pmatrix}
H^+ \\
\frac{1}{\sqrt{2}}\big(\varphi_2^0 + i a^0\big)
\end{pmatrix}\,,
\end{equation}
where $v^2 = (246.2\,\text{GeV})^2$. The unphysical ``would-be''
Goldstone boson fields are denoted by $G^+$ and $G^0$. The physical
charged Higgs bosons $H^\pm$ have a squared mass given by
\begin{equation}
 M_{H^\pm}^2 = Y_2 + \frac{1}{2}Z_3 v^2 \,.
\end{equation}
In the presence of CP-violation, all three neutral Higgs bosons
$\varphi_1^0$, $\varphi_2^0$, and $a^0$ mix. The corresponding
symmetric squared-mass matrix $\mathcal{M}^2$ is given by
\begin{equation}\label{eq:neutralMassMtx}
 \frac{\mathcal{M}^2}{v^2} =
\begin{pmatrix}
Z_1 & \text{Re}(Z_6) & -\text{Im}(Z_6) \\[0.5ex]
\text{Re}(Z_6) & Y_2/v^2+\frac{1}{2}Z^+_{345} & -\frac{1}{2}\text{Im}(Z_5 ) \\[0.5ex]
 -\text{Im}(Z_6) & -\frac{1}{2}\text{Im}(Z_5 ) &  Y_2/v^2+\frac{1}{2}Z^-_{345}
\end{pmatrix}\,,
\end{equation}
where $Z_{345}^\pm = Z_3 + Z_4 \pm \text{Re}(Z_5)$. This mass matrix
is diagonalized by a special orthogonal transformation that can be
parameterized as
\begin{equation}
\label{eq:HBtoMassRot}
\begin{pmatrix}
h_1 \\ h_2 \\ h_3
\end{pmatrix} = \begin{pmatrix}
q_{11} & \text{Re}(q_{12}) & \text{Im}(q_{12})\\
q_{21} & \text{Re}(q_{22}) & \text{Im}(q_{22})\\
q_{31} & \text{Re}(q_{32}) & \text{Im}(q_{32})
\end{pmatrix}
\begin{pmatrix}
\varphi_1^0 \\
\varphi_2^0 \\
a^0
\end{pmatrix}\,.
\end{equation}
We denote the masses of the three neutral mass-eigenstate Higgs bosons
$h_k$ by $M_{h_k}$, for $k=1,2,3$. The elements of the rotation matrix
in Eq.~\eqref{eq:HBtoMassRot} can be parameterized by three angles
\begin{equation}
\begin{pmatrix}
q_{11} & \text{Re}(q_{12}) & \text{Im}(q_{12})\\
q_{21} & \text{Re}(q_{22}) & \text{Im}(q_{22})\\
q_{31} & \text{Re}(q_{32}) & \text{Im}(q_{32})
\end{pmatrix}
=
\begin{pmatrix}
c_{12} c_{13}&- s_{12} c_{23} - c_{12} s_{13} s_{23}& s_{12} s_{23}-c_{12}s_{13}c_{23}\\
s_{12} c_{13}& c_{12} c_{23} - s_{12} s_{13} s_{23} &-c_{12} s_{23}-s_{12}s_{13}c_{23}\\
s_{13} & c_{13} s_{23} & c_{13} c_{23}
\end{pmatrix}\,,
\end{equation}
where $s_{ij}$ ($c_{ij}$) is $\sin\theta_{ij}$
($\cos\theta_{ij}$). This parameterization is used in the
\texttt{python} code (see Sec.~\ref{sec:conclusions}), such that the
orthogonality of the rotation matrix is manifest and no spurious
entries remain.

The Yukawa interactions in the mass-eigenstate basis for the Higgs
bosons and the fermions are given by~\cite{Altmannshofer:2024mbj} (see
also Ref.~\cite{Boto:2020wyf})
\begin{equation}\label{eq:Yukawa}
\begin{split}
\mathcal{L}_\text{Yuk}
 & \supset
  - \sum_k \sum_{ij} h_k \bigg[
  \bar{u}_{L,i}
  \left( \frac{m_{u_i}}{v} \delta_{ij} q_{k1} + \rho_{u,ij} q^*_{k2} \right) u_{R,j} \\
& \hspace{6em} + \sum_{f=d,e}
  \bar{f}_{L,i}
  \left( \frac{m_{f_i}}{v} \delta_{ij} q_{k1} + \rho_{f,ij}^\dagger q_{k2} \right) f_{R,j}
  \bigg] + \text{c.c.}  \\
& \quad 
- \sqrt{2} H^+ \big[ \bar{u}_{L,i} \big(V \rho_d^\dagger \big)_{ij} d_{R,j}
                    - \bar{u}_{R,i} \big(\rho_u^\dagger V \big)_{ij} d_{L,j} \big]
- \sqrt{2} H^+ \bar{\nu}_{L,i} \rho_{\ell,ij}^\dagger e_{R,j} + \text{c.c.} \,,
\end{split}
\end{equation}
Here, $d_L$, $u_L$, and $e_L$ ($d_R$, $u_R$, and $e_R$) denote the
left-(right-)handed up-quark, down-quark, and charged lepton fields,
respectively. The left-handed neutrinos are denoted with $\nu_L$. We
do not display the interactions of the unphysical Goldstone bosons,
but they were included in our calculations to obtain gauge-independent
results. The $3\times 3$ matrices ${\rho}_f$ contain general complex
entries that comprise flavor and $CP$ violation. The charged Higgs
interactions with quarks contain the Cabibbo-Kobayashi-Maskawa (CKM)
quark mixing matrix $V$.

In practice, we performed all calculations using the background field
gauge method in the electroweak sector~\cite{Denner:1994xt}, since
this leads to a drastic reduction in the complexity of the
calculation. In the unbroken phase, we split all gauge fields into
quantum fields and background fields (the latter denoted by a hat). We
also perform a similar splitting for the field $H_1$ in the Higgs
basis, i.e. we have
\begin{equation}
  \hat H_1 =
\begin{pmatrix}
  \hat G^+\\
  \frac{1}{\sqrt{2}} \big( v + \hat \phi_1^0 + i \hat G^0 \big)
\end{pmatrix}\,, \qquad
  H_1 =
\begin{pmatrix}
  G^+\\
  \frac{1}{\sqrt{2}} \big( \phi_1^0 + iG^0 \big)
\end{pmatrix}\,.
\end{equation}
We then add the following gauge-fixing term to the Lagrangian,
\begin{equation}
\begin{split}
  {\mathcal L}_\text{gf}
& = - \frac{1}{2\xi}
      \bigg[ \big( \delta^{ac} \partial_\mu
                   + g_2 \epsilon^{abc} \hat W_\mu^b \big) W^{c,\mu}
             -ig_2 \xi \frac{1}{2}
             \big( \hat H_{1,i}^\dagger \sigma_{ij}^a H_{1,j} 
                   - H_{1,i}^\dagger \sigma_{ij}^a \hat H_{1,j}\big) \bigg]^2 \\
& \quad - \frac{1}{2\xi}
          \bigg[ \partial_\mu B^\mu 
                 +ig_1 \xi \frac{1}{2}
                 \big( \hat H_{1,i}^\dagger H_{1,i} 
                       - H_{1,i}^\dagger \hat H_{1,i}\big) \bigg]^2 \,.
\end{split}
\end{equation}
Here, $g_1$ and $g_2$ are the gauge couplings associated with the
hypercharge $U(1)_Y$ gauge boson $B_\mu$ and the triplet of weak $SU(2)_L$
gauge bosons $W_\mu^a$, respectively, and $\xi$ is an arbitrary gauge
parameter, chosen to be the same for all weak gauge bosons, to avoid
tree-level mixing between the photon and the $Z$ boson (see
Ref.~\cite{Denner:1994xt} for details). Moreover,
$s_w \equiv \sin\theta_w$ denotes the sine of the weak mixing angle,
and $c_w \equiv \cos\theta_w$. The Feynman rules are obtained by
expressing all fields in the broken phase and rotating to mass
eigenstates. The ghost Lagrangian is constructed in the usual
way~\cite{Denner:1994xt}. For our calculation, only the Feynman rules
with a single background photon field are needed. In our conventions,
$\hat A^\mu = c_w \hat B^\mu - s_w \hat W^{3,\mu}$.




\section{Renormalization-group analysis}\label{sec:rg}


Our goal in this work is to calculate the leading contributions to the
coefficients of the low-energy effective Lagrangian, given in
Eq.~\eqref{L:dipole:3f} below, that contributes to hadronic, atomic,
and molecular EDMs. This low-energy Lagrangian is defined at a
renormalization scale of $\mu \sim 2\,$GeV, and comprises the up-,
down-, and strange-quark EDM and chromo-EDM operators, CP-odd
semileptonic operators, as well as the Weinberg three-gluon
operator. We restrict ourselves to flavor diagonal dipole operators,
but we do include flavor off-diagonal couplings in the 2HDM. The
hadronic matrix elements and contributions to observables are
discussed in several reviews, see, e.g., Refs.~\cite{Engel:2013lsa,
  FlavourLatticeAveragingGroupFLAG:2024oxs}.

\bigskip

We do not assume that the Higgs bosons have masses far above the weak
scale. We will therefore match the full 2HDM onto an effective
five-flavor Lagrangian at the weak scale, and our results are valid
for any Higgs masses at or above the weak scale.\footnote{For
  Higgs-boson masses that are very much heavier than the weak scale
  one could alternatively match to an intermediate effective theory
  (EFT), such as SMEFT~\cite{Buchmuller:1985jz, Grzadkowski:2010es} or
  HEFT~\cite{Feruglio:1992wf}, and sum the corresponding large
  logarithms of ratios of weak-scale and Higgs masses. This is not
  expected to lead to large corrections. Note, also, that all
  contributions to CP-odd observables are suppressed by an inverse
  quadratic power of the (large) Higgs masses.}

It is important to note that QCD effects are large for all hadronic
EDMs~\cite{Braaten:1990gq, Braaten:1990zt, Degrassi:2005zd,
  Hisano:2012cc, Brod:2018pli, Brod:2023wsh, Kumar:2024yuu}, and a
realistic estimate requires the systematic summation, using
RG-improved perturbation theory, of all large logarithms of ratios of
weak-scale masses and the hadronic scale of 2\,GeV. The details of the
RG evolution depend on the fermion masses appearing in the
diagrams. Accordingly, the matching conditions at the weak scale will
need to be evaluated at tree-level, at one-loop, and at two-loop,
depending on which combination of masses and couplings of the involved
fermions give the numerically leading contributions. As a guide to our
calculation, we now discuss the different cases.

First, we point out that all leading flavor-diagonal contributions to
CP-odd observables must involve the exchange of virtual Higgs bosons.
Without the additional phases in the Higgs and Yukawa sectors,
contributions to electric dipole moments arise from either the QCD
$\theta$ term, or the phase of the CKM matrix. Here, we work under the
assumption that the $\theta$ term is negligibly small. The effect of
the CKM phase on hadronic and leptonic EDM is strongly suppressed by
the associated flavor violation (see Refs.~\cite{Seng:2014lea,
  Yamaguchi:2020eub, Ema:2022yra} for recent estimates).

\begin{figure}[t]
  \centering
  \includegraphics[width=12em]{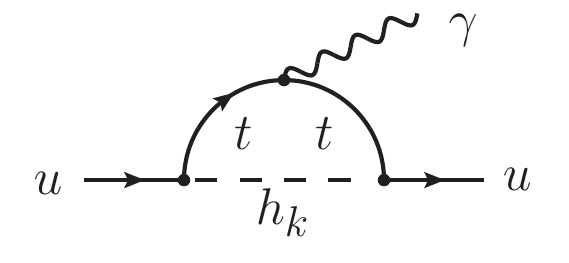}~~~~
  \includegraphics[width=12em]{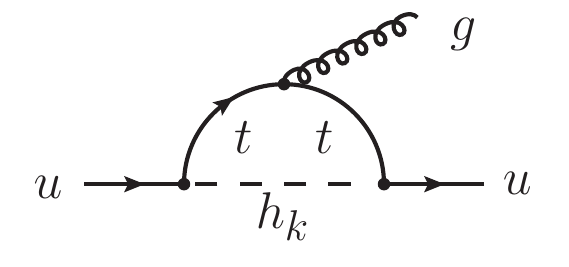}~~~~
  \includegraphics[width=12em]{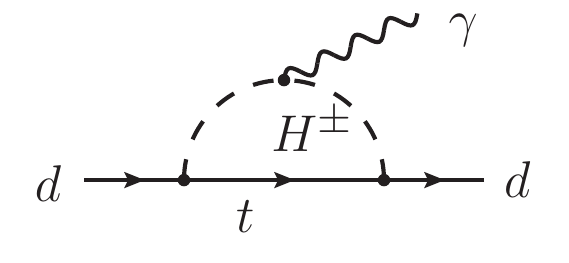}
  \caption{Representative Feynman diagrams for the one-loop
    contributions to the EDM (left panel) and chromo-EDM (center
    panel) of the up quark, and to the EDM of the down quark (right
    panel). Solid lines denote quarks, and dashed lines denote Higgs
    bosons (labelled by $h_k$ and $H^\pm$). 
    \label{fig:edm:1loop}}
\end{figure}

The leading contributions to the quark EDMs and CEDMs then start at
one-loop order, via the exchange of virtual charged or neutral Higgs
bosons (see Fig.~\ref{fig:edm:1loop}). If any of the fermions in these
one-loop diagrams have masses below the weak scale, the result will be
dominated by the logarithm of the large ratio of fermion to Higgs
mass. In this case, we use the RG to calculate the coefficient of this
logarithm via the mixing of an appropriate four-fermion operator into
the dipole, hence only the tree-level initial condition of that
four-fermion operator needs to be calculated. Neglecting the
non-logarithmic part of the diagram results in a much smaller error
than neglecting the higher-order QCD corrections. If, on the other
hand, all particles in the one-loop dipole diagram are heavy (masses
of the order of the weak scale), the logarithm is small and the full
result of the diagram is needed to calculate the initial
condition. This is the case only for the down-quark dipoles with an
internal charged Higgs and top quark, and for the up-quark dipoles
with neutral Higgs bosons that have flavor-violating couplings to top
quarks.

It has been pointed out long ago by Weinberg~\cite{Weinberg:1989dx}
and Dicus~\cite{Dicus:1989va}, and by Barr and Zee~\cite{Barr:1990vd}
that, if there is a large hierarchy between the Yukawa couplings, the
numerically leading contribution to the dipole operators might in fact
arise from two-loop diagrams. Moreover, the contributions to the gluon
chromo-EDM (the ``Weinberg operator''~\cite{Weinberg:1989dx}) always
start at two-loop order (see Fig.~\ref{fig:WBO} below). To be fully
general, we do not assume any hierarchy between the Yukawa couplings
(although, most likely, such a hierarchy would be realized with
realistic values for the couplings). Accordingly, we calculate all
leading two-loop contributions, i.e., those that probe products of
couplings that are not appearing in the one-loop contributions. (In
the latter case, the two-loop diagrams would just constitute a small
radiative correction to the leading one-loop result.)

Again, the full two-loop diagrams should only be used if no large
logarithms appear in the result (this is the case, for instance, if
the only virtual fermion in the loop is the top quark). The reason is
-- as in the case of one-loop contributions -- that the large size of
the QCD corrections requires the consistent summation of all large
logarithms. If all logarithms are small (i.e. the loop function
depends only on particle masses of the order of the electroweak
scale), the full two-loop result is used as an initial condition for
the EDM and chromo-EDM operators and the Weinberg operator, which then
mix among each other when running to the hadronic scale. For instance,
the loop function arising from diagrams with virtual bottom and charm
quarks frequently exhibits a quadratic large logarithm of the form
$\log^2(m_q/M_h)$, $q=b,c$. In this case, only the tree-level initial
condition of the appropriate four-quark operator is needed. In fact,
using the full two-loop result would lead to a large error (see
Ref.~\cite{Brod:2013cka} and the discussion in
Sec.~\ref{sec:light}). In some cases, the two-loop result exhibits a
{\em linear} large logarithm. In our context, this logarithm can be
generated either via a threshold correction at a heavy-quark scale
(this happens, for instance, for the bottom-quark contribution to the
Weinberg operator), or via an one-loop initial condition and
subsequent RG evolution to the hadronic scale (examples are fermionic
Barr-Zee diagrams involving the exchange of virtual charged Higgs
bosons, discussed in Sec.~\ref{sec:edm}). In some cases, even QED
effects have to be calculated to capture the leading effects. This is
also discussed in Sec.~\ref{sec:light}.

\bigskip

In the following sections
we define the effective five-, four-, and three-flavor Lagrangians,
and discuss the renormalization group evolution between the weak scale
and the low energy flavor thresholds, as well as the relevant
threshold corrections.

\subsection{Effective five- and four-flavor Lagrangians}\label{sec:fivefour}

The QCD corrections to the hadronic four-quark and dipole operators
are large. Accordingly, we using RG-improved perturbation theory to
sum all leading logarithms. The set of hadronic operators that
contributes to the three-flavor Lagrangian via running and / or
matching, and is closed under QCD RG, is
\begin{equation} \label{eq:Leff5}
\begin{split}
  {\mathcal L}_\text{eff,5f}
= - \frac{1}{v^2} 
  & \sum_{q\neq q'}
    \bigg[ \sum_{i=1,2} C_i^{qq'}  Q_i^{qq'} + \frac{1}{2} \sum_{i=3,4} C_i^{qq'}  Q_i^{qq'} \bigg]\\
  - \frac{1}{v^2}
  & \sum_{q} \sum_{i=1}^4 
    \bigg[   C_i^q   Q_i^q \bigg]
  - \frac{1}{v^2} C_w  Q_w \\
  - \frac{1}{v^3}
  & \sum_{q,\ell}
    \bigg[   C_1^{\ell q}   Q_1^{\ell q} + C_1^{q\ell}   Q_1^{q\ell}
           + C_2^{\ell q}   Q_2^{\ell q} \bigg]
  - \frac{1}{v^3} C_5^e Q_5^e \,.
\end{split}
\end{equation}
The summation runs over all leptons and quarks with masses below the
weak scale ($\ell = e,\mu,\tau$; $q,q' = u,d,s,c,b$). The four-flavour
effective Lagrangian can simply be obtained by excluding the bottom
quark from the sum. The four-quark operators are defined as
\begin{align}
\label{eq:op:qq':1:2}
Q_1^{qq'} & = (\bar q q) \, (\bar q' \, i \gamma_5 q') \,,
&
Q_2^{qq'} & = (\bar q \, T^a  q) \, (\bar q' \, i \gamma_5 T^a q') \,,\\[0.5em]
\label{eq:op:qq':3:4}
Q_3^{qq'} & = \frac{1}{2}\epsilon^{\mu\nu\rho\sigma} (\bar q \sigma_{\mu\nu} q) \, (\bar q' \, \sigma_{\rho\sigma} q') \,,
&
Q_4^{qq'} & = \frac{1}{2}\epsilon^{\mu\nu\rho\sigma} (\bar q \sigma_{\mu\nu} T^a q) \, (\bar q' \sigma_{\rho\sigma} T^a q') \,,\\[0.5em]
\label{eq:op:q}
Q_1^q & = (\bar q q) \, (\bar q \, i \gamma_5 q) \,,
&
Q_2^q & = \frac{1}{2}\epsilon^{\mu\nu\rho\sigma}(\bar q \sigma_{\mu\nu} q) \, (\bar q \, \sigma_{\rho\sigma} q) \,,
\end{align}
and the semileptonic operators are
\begin{align}\label{eq:op:semileptonic}
Q_1^{\ell q} & = m_q (\bar \ell \ell) \, (\bar q \, i \gamma_5 q) \,,
&
Q_1^{q \ell} & = m_q (\bar \ell \, i \gamma_5 \ell) \, (\bar q q) \,,
&
Q_2^{\ell q} & = \frac{m_q }{2}\epsilon^{\mu\nu\rho\sigma}
                (\bar \ell \sigma_{\mu\nu} \ell) \, (\bar q \, \sigma_{\rho\sigma} q) \,.
\end{align}
Here, $\epsilon^{\mu\nu\rho\sigma}$ is the completely antisymmetric
Levi-Civita tensor, with sign convention $\epsilon_{0123} = 1$, and we
have defined $\sigma^{\mu\nu} = \tfrac{i}{2} [\gamma^\mu,
  \gamma^\nu]$. Note that we have included an explicit quark mass
factor in the definition of the semileptonic operators, and a
corresponding additional normalization factor of $1/v$ in the
Lagrangian~\eqref{eq:Leff5}, to avoid ratios of quark and lepton
masses in the anomalous dimensions of these operators,
Eq.~\eqref{eq:sl:adm}. Finally, we define the dimension-seven
operator,
\begin{align}\label{eq:op:eeGG}
Q_5^e & = \frac{\alpha_s}{12\pi} \bar e i \gamma_5 e \, G^a_{\mu \nu} G^{a,\mu\nu} \,.
\end{align}
It induces a CP-violating electron-nucleon coupling, and captures the
effects of $Q_1^{b \ell}$ and $Q_1^{c \ell}$ at low energies, see
Eq.~\eqref{eq:th:C5e}.

The dipole operators are normalized in a way that gives a convenient
form of the anomalous dimension. We use the following definitions,
\begin{align}
Q_3^q & = \frac{eQ_q}{2} \frac{m_q}{g_s^2} \,
         (\bar q \sigma^{\mu \nu} i\gamma_5 q) \, F_{\mu\nu} \,, \label{eq:Q3q} \\
Q_4^q & = -\frac{1}{2} \, \frac{m_q}{g_s} \,
         (\bar q \sigma^{\mu \nu} i\gamma_5 T^a q) \, G^a_{\mu\nu} \,, \label{eq:Q4q} \\
Q_w & = -\frac{1}{3g_s} f^{abc} \,
         G_{\mu \sigma}^a G_{\nu}^{b, \sigma} \widetilde G^{c, \mu \nu} \label{eq:Qw} \,.
\end{align}
Here, $F_{\mu\nu} = \partial_\mu A_\nu - \partial_\nu A_\mu$ and
$G_{\mu\nu}^a = \partial_\mu G_\nu^a - \partial_\nu G_\mu^a + g_s
f^{abc} G_\mu^b G_\nu^c$ are the field-strength tensors for QED and
QCD, respectively. The dual gluon field strength tensor is defined as
$\tilde{G}^{a,\mu\nu} = \frac{1}{2} \epsilon^{\mu\nu\alpha\beta}
G_{\alpha\beta}^a$, with $\epsilon^{0123} = -1$. In our convention,
the covariant derivative acting on quark fields is given by
\begin{equation}
  D_\mu = \partial_u - ig_s T^a G_\mu^a + ieQ_q A_\mu \,.
  \label{eq:quarkcovariantderivative}
\end{equation}
$Q_f$ is the charge of fermion $f$ in units of the positron charge
$e$. The basis of all CP-odd operators is closed under the RG flow of
both QCD and QED as both interactions are CP conserving.

\subsection{RG evolution}\label{sec:rge}

We define the anomalous dimensions via the RG equation for the Wilson
coefficients,
\begin{equation}
  \frac{d C_i}{d \log \mu} = \sum_j C_j \gamma_{ji} \,.
\end{equation}
The explicit expressions for the QCD anomalous dimensions are
collected in this section. (QED effects are discussed in
Sec.~\ref{sec:light}.) We keep only the leading contributions in the
strong coupling constant, expanding
$\gamma = (\alpha_s/4\pi) \gamma^{(0)} + \ldots\,$.

The anomalous dimensions for the dipole and hadronic operators can be
taken from Ref.~\cite{Brod:2018pli} (the original references for the
one-loop results are~\cite{Morozov:1984goy, Braaten:1990gq,
  Braaten:1990zt, Degrassi:2005zd, Hisano:2012cc}):
\begin{align}
\gamma^{(0)}_{q\to q} &=
\begin{pmatrix}
-10 &   -\frac{1}{6} &                               4 &    4                            \\
 40 &   \frac{34}{3} &                           - 112 & - 16                            \\
  0 &              0 & -\frac{34}{3} + \frac{4}{3} N_f &    0                            \\
  0 &              0 &  \frac{32}{3}                   & -\frac{38}{3} + \frac{4}{3} N_f \\
\end{pmatrix}\,,\\
\gamma^{(0)}_{q q' \to q} &=
\begin{pmatrix}
0 & 0 & 0 & 0\\
0 & 0 & 0 & 0\\
0 & 0 & 0 & 0\\
0 & 0 & 0 & 0\\
0 & 0 &  - 48 \frac{Q_{q'}}{Q_q} \frac{m_{q'}}{m_q}& 0\\
0 & 0 &  0& - 8 \frac{m_{q'}}{m_q}
\end{pmatrix}\,, \label{eq:adm_qqprime_q}\\
\gamma^{(0)}_{qq' \to qq'} &=
\begin{pmatrix}
 -16& 0& 0& 0& 0& - 2\\
 0&2& 0& 0& - \frac{4}{9}& - \frac{5}{6}\\
 0& 0& -16& 0& 0& - 2\\
 0& 0& 0&2& - \frac{4}{9}& - \frac{5}{6}\\
 0& - 48& 0& - 48&\frac{16}{3}& 0\\
 - \frac{32}{3}& - 20& - \frac{32}{3}& - 20& 0&-\frac{38}{3}
\end{pmatrix}\,,\\
\gamma^{(0)}_{W \to q} &=
\begin{pmatrix}
  0 & 0 & 0 & 6
\end{pmatrix}\,,\\
\gamma^{(0)}_{W\to W} &=  -8+\frac{8}{3}N_f\,.
\end{align}
Here, $N_f$ denotes the number of active quark flavors. In our matrix
notation, we used the following ordering of the Wilson coefficients:
\begin{equation}\label{eq:WC:ordering}
\begin{split}
q = \{
&C_1^{q},\,C_2^{q},\,C_3^{q},\,C_4^{q}\} \,, \\
qq' = \{
&C_1^{qq'},\, C_2^{qq'},\, C_1^{q'q},\, C_2^{q'q},\, C_3^{qq'},\, C_4^{qq'}\} \,.
\end{split}
\end{equation}

The semileptonic operators evolve like currents under QCD, see
Ref.~\cite{Brod:2023wsh}. Using the results of
Refs.~\cite{Brod:2018pli, Bishara:2018vix}, we see that the scalar
operators in Eq.~\eqref{eq:op:semileptonic} do not mix into the dipole
operators Eqs.~\eqref{eq:Q3q} and~\eqref{eq:Q4q}. The tensor operators
in Eq.~\eqref{eq:op:semileptonic}, on the other hand, do mix into the
dipole operators. Their leading non-zero initial condition arise at
one-loop for the tensor operators with up-type quarks (see
Eq.~\eqref{eq:C2lu}, \eqref{eq:C2ld} below). This corresponds to the
appearance of a large linear logarithm in some of the two-loop
diagrams. Using the ordering
\begin{equation}
\begin{split}
\ell q = \{
&C_1^{\ell q},\, C_1^{q \ell},\, C_2^{\ell q}\} \,,
\end{split}
\end{equation}
we find the following anomalous dimensions~\cite{Brod:2023wsh})
\begin{align}
\gamma^{(0)}_{\ell q \to \ell q} &=
\begin{pmatrix}
 0& 0& 0\\
 0& 0& 0\\
 0& 0& \frac{32}{3}
\end{pmatrix}\,,
& \gamma^{(0)}_{\ell q \to q} &=
\begin{pmatrix}
0 & 0 & 0 & 0\\
0 & 0 & 0 & 0\\
0 & 0 &  - 16 \frac{Q_{\ell}}{Q_q} \frac{m_{\ell}}{v} & 0
\end{pmatrix}\,, \label{eq:sl:adm}\\
\gamma^{(0)}_{Q_e^5 \to e q} &=
\begin{pmatrix}
  0& 0 & 0
\end{pmatrix}\,,
&\gamma^{(0)}_{Q_e^5 \to Q_e^5} &= 0 \,. \label{eq:Qe5:adm}
\end{align}
The operator $Q_e^5$ mixes into the semileptonic operator
$Q_1^{q \ell}$ at order $\alpha_s^2$. The effect is numerically tiny,
and we will neglect it. A detailed discussion of large logarithms
arising from light fermion masses, including QED effects, is given in
Sec.~\ref{sec:light}.

Finite contributions to the Wilson coefficients arise when crossing a
heavy-flavor threshold. At the bottom-quark threshold, the
coefficients of $Q_5^e$ and the Weinberg operator receive a finite
matching correction at leading order~\cite{Braaten:1990gq,
  Braaten:1990zt, Bishara:2018vix}
\begin{align}
  C_5^e|_{n_f=4} (\mu_b) & = C_5^e|_{n_f=5} (\mu_b) - C_1^{be} |_{n_f=5} (\mu_b) \,,
                           \label{eq:th:C5e} \\
  C_{w}|_{n_f=4} (\mu_b) & = C_{w}|_{n_f=5} (\mu_b)
                           - \frac{\alpha_s^{(5)}}{8\pi} C_{4}^{b} |_{n_f=5} (\mu_b) \,.
                           \label{eq:th:wbo}
\end{align}
An analogous contribution arises at the charm-quark threshold.

\subsection{Effective three-flavor Lagrangian and dipole coefficients}\label{sec:dipole}

After evolving all Wilson coefficients to the hadronic scale
$\mu_\text{had} = 2\,$GeV, we slightly change the coefficients of the
(chromo-)electric dipole operators to match the form used in
evaluating the corresponding hadronic matrix elements, see
Ref.~\cite{Engel:2013lsa}. We use the following CP-odd,
flavor-conserving Lagrangian to parameterize the contributions to
quark and gluon (chromo-)EDMs:
\begin{equation} \label{L:dipole:3f}
\begin{split}
\mathcal L_\text{dipole,3f} & =
 - \sum_{q=u,d,s} \frac{d_q}{2} (\bar q \sigma^{\mu\nu} i\gamma_5 q) F_{\mu\nu}
 - \sum_{q=u,d,s} \frac{g_s\tilde d_q}{2} (\bar q \sigma^{\mu\nu} i\gamma_5 T^a q) G^a_{\mu\nu} \\
& \quad 
 + \frac{d_w}{3} f^{abc} \, G_{\mu \sigma}^a G_{\nu}^{b, \sigma} \widetilde G^{c, \mu \nu}
 + \ldots \,.
\end{split}
\end{equation}
The ellipsis denotes the remaining, unchanged contributions in the
three-flavor effective theory. (In particular, we do not change the
definitions of the four-fermion operators.)  The remaining operators
have been defined in Eqs.~\eqref{eq:op:qq':1:2}-\eqref{eq:op:eeGG};
the summation here is over the light quarks only, $q=u,d,s$. The
relation between the previously defined Wilson coefficients and the
dipole moments for the hadronic EDMs is
\begin{align}
  d_q(\mu_\text{had})
 & = \frac{1}{v^2} \frac{Q_qe}{4\pi\alpha_s}
     m_q C_3^q(\mu_\text{had}) \,, \\
     \tilde d_q(\mu_\text{had})
 & = - \frac{1}{v^2} \frac{1}{4\pi\alpha_s}
     m_q C_4^q(\mu_\text{had}) \,, \\
     d_w(\mu_\text{had})
 & = \frac{1}{v^2} \frac{1}{g_s} C_w^q(\mu_\text{had}) \,. \label{eq:dw}
\end{align}
The strong coupling constant and all quark masses should be evaluated
at $\mu_\text{had} = 2\,$GeV.

\section{Weak-scale matching}\label{sec:match}

In this section, we provide the explicit results for the matching
conditions of the 2HDM described in Sec.~\ref{sec:2hdm} onto the
five-flavor effective Hamiltonian~\eqref{eq:Leff5}. These matching
conditions are the main results of this work. The Wilson coefficients
are subsequently evolved to the hadronic scale using the RG equations,
as explained in Sec.~\ref{sec:rge}.

All Feynman diagrams have been calculated using the package
\texttt{MaRTIn}~\cite{Brod:2024zaz}, based on
\texttt{FORM}~\cite{Vermaseren:2000nd} and implementing the algorithms
developed in Refs.~\cite{Davydychev:1992mt, Bobeth:1999mk}. The
diagrams have been generated using
\texttt{qgraf}~\cite{Nogueira:1991ex}, and typeset in latex using
\texttt{jaxodraw}~\cite{Binosi:2003yf}, based on
\texttt{axodraw}~\cite{Vermaseren:1994je}. We have calculated all
diagrams in generalized $R_\xi$ gauge for the photon and all weak
gauge bosons~\cite{Altmannshofer:2024mbj}, and explicitly verified the
gauge-parameter independence of all our final results. The Feynman
rules for the unphysical Goldstone bosons are taken from
Ref.~\cite{Brod:2019bro}. As a further check, all results have been
obtained by at least two independent calculations.

The calculation of the two-loop diagrams with internal closed fermion
loops (Fig.~\ref{fig:BZ:fermion}, left panel) involves the evaluation
of fermion traces containing one or more powers of $\gamma_5$. For
reasons of algebraic consistency, we employ the 't~Hooft-Veltman
scheme for $\gamma_5$, involving mixed commutation and anticommutation
relations (see Ref.~\cite{tHooft:1972tcz, Breitenlohner:1977hr,
  Collins:1984xc} for details). Our calculations are consistent with
the available results in the literature, after the proper inclusion of
finite counterterm contributions, as described in
Ref.~\cite{Altmannshofer:2024mbj}.

In all our results, we use the approximation of a unit CKM matrix. For
all tree-level results (contained in Sec.~\ref{sec:four:quark} and
Sec.~\ref{sec:semi:leptonic}) it is straightforward to include the
effects of the full CKM matrix $V$ by replacing
$\rho_d \to \rho_d V^\dagger$ and $\rho_u \to V^\dagger \rho_u$ in the
charged-Higgs contributions, cf. Eq.~\eqref{eq:Yukawa}. In some of the
loop-level contributions, this replacement (and the corresponding
modifications of the charged weak currents) would imply additional
quark mass dependence in the loop functions, so that not all those
results can be easily generalized for a full CKM matrix. The terms that
are neglected in our approximation are suppressed by small quark-mass
to gauge- and Higgs-boson mass ratios and / or small off-diagonal CKM
matrix elements. They are not expected to have a sizeable effects on
the phenomenology.

We present the fermionic contributions to the quark electric dipole
moments in Sec.~\ref{sec:edm}. The contributions of light fermions
that are enhanced by a large logarithm require a separate treatment,
presented in Sec.~\ref{sec:light}. The bosonic contributions to the
quark EDMs can be found in Sec.~\ref{sec:bosonic}. We give only the
results for the up, down, and strange EDMs (the charm and bottom EDMs
would receive some additional contributions that we did not
calculate). The quark chromo-EDMs and the contribution to the Weinberg
operator are presented in Sec.~\ref{sec:cedm} and Sec.~\ref{sec:wbo},
respectively. Finally, in Sec.~\ref{sec:four:quark} we present the
contributions to the CP-violating four-quark operators, and in
Sec.~\ref{sec:semi:leptonic} the contributions to the semi-leptonic
operators. All results presented in this section are new in the sense
that, to the best of our knowledge, they have never been calculated in
fully analytic form and for arbitrary phases in both the Higgs
potential and the Yukawa couplings.

\subsection{Fermionic contributions to the quark EDMs}\label{sec:edm}

\begin{figure}[t]
  \centering
  \includegraphics[width=12em]{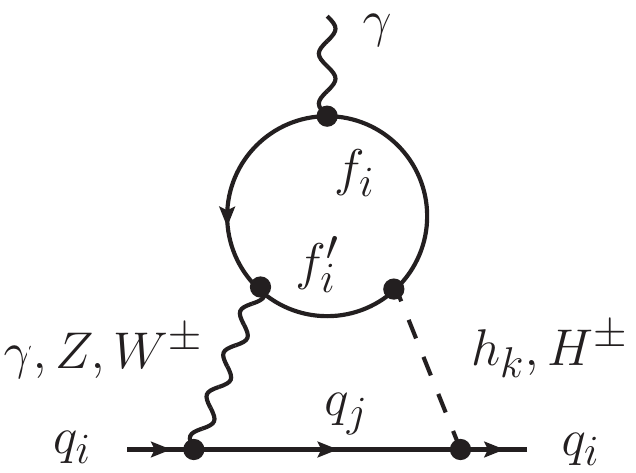}~~~
  \includegraphics[width=12em]{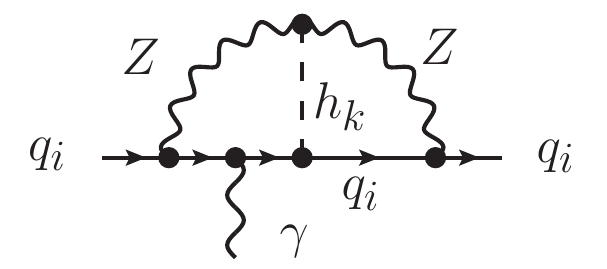}
  \caption{Representative two-loop diagrams contributing to the
    light-quark EDM. \label{fig:BZ:fermion}}
\end{figure}

We split the initial conditions at the weak scale into their one-loop
and two-loop contributions; they should be summed to obtain the total
result. Below the weak scale, additional contributions (corresponding
to certain light-fermion loops) are generated by the RG evolution, see
Sec.~\ref{sec:light}.

The only non-zero contributions at one-loop involve diagrams with
internal top quarks. We find
\begin{align}
  C_3^{u_i, \text{one-loop}} & = - \frac{\alpha_s}{4\pi} \frac{m_t}{m_{u_i}}
               \sum_k \frac{v^2}{M_{h_k}^2} g_1(x_{t h_k})
               \text{Im} \big\{\rho_{u,3i}^* \rho_{u,i3}^* q_{k2}^2 \big\} \,, \label{eq:C34u} \\
  C_3^{d_i, \text{one-loop}} & = - \frac{\alpha_s}{4\pi} \frac{m_t}{m_{d_i}} \frac{v^2}{M_{H^\pm}^2}
                 \big[ 4 g_1(x_{t H^\pm}) + 3 g_2(x_{t H^\pm}) \big]
                 \text{Im}\big\{ \rho^*_{u,i3} \rho_{d,i3} \big\} \,. \label{eq:C3d}
\end{align}
Here and below, we define the squared mass ratios
$x_{ab} \equiv M_{a}^2/M_{b}^2$. The one-loop functions are defined by
\begin{equation}\label{eq:g12}
  g_1(x) = \frac{x - 3}{2(x-1)^2} + \frac{1}{(x-1)^3} \log x \,, \qquad
  g_2(x) = \frac{x + 1}{(x-1)^2} - \frac{2x}{(x-1)^3} \log x \,.
\end{equation}

It is convenient to split the two-loop contribution into several terms
as follows. For the up-quark EDM, we have
\begin{equation}
\begin{split}
  C_3^{u, \text{two-loop}}
& = C_3^{u, t h}
  + C_3^{u, t H^\pm}  + C_3^{u, c H^\pm} \\
& \quad + C_3^{u, hZ} + C_3^{u, \text{kin.}}
  + C_3^{u, H^\pm \gamma} + C_3^{u, H^\pm Z} + C_3^{u, H^\pm W} \,, \\
\end{split}
\end{equation}
and for the down- and strange-quarks EDMs, we have
\begin{equation}
\begin{split}
  C_3^{d_i, \text{two-loop}}
& = C_3^{d_i, t h}
  + C_3^{d_i, t H^\pm} + \sum_j C_3^{d_i, \ell_j H^\pm} \\
& \quad + C_3^{d_i, hZ} + C_3^{d_i, \text{kin.}}
  + C_3^{d_i, H^\pm \gamma} + C_3^{d_i, H^\pm Z} + C_3^{d_i, H^\pm W} \,.
\end{split}
\end{equation}
All the individual terms in these sums are separately gauge
invariant. They correspond to Barr-Zee diagrams with internal
top-quark loops and neutral Higgs bosons ($C_3^{u, t h}$ and
$C_3^{d_i, t h}$), Barr-Zee diagrams with internal top-quark loops and
charged Higgs bosons ($C_3^{u, t H^\pm}$ and $C_3^{d_i, t H^\pm}$),
Barr-Zee diagrams with internal light fermion loops and charged Higgs
bosons ($C_3^{u, c H^\pm}$ and $C_3^{d_i, \ell_j H^\pm}$), and
non-Barr-Zee diagrams with internal $Z$ and neutral Higgs bosons
($C_3^{u, hZ}$ and $C_3^{d_i, hZ}$). The remaining diagrams involve
couplings from the Higgs kinetic terms ($C_3^{u, \text{kin.}}$ and
$C_3^{d_i, \text{kin.}}$), as well as couplings from the Higgs
potential ($C_3^{u, H^\pm V}$ and $C_3^{d_i, H^\pm V}$), with
$V = \gamma, Z, W$. The explicit expressions for the various terms are
given below.

The contributions of Barr-Zee diagrams with internal top-quark loops
and neutral Higgs bosons (see Fig.~\ref{fig:BZ:fermion}, left panel)
can be obtained by simple replacements from the results in
Ref.~\cite{Altmannshofer:2024mbj}. We find for the down and strange
EDMs ($d_1 = d$, $d_2 = s$)
\begin{equation}
\begin{split}\label{eq:C3di}
  & C_3^{d_i,t h}
  = \frac{3\alpha\alpha_s Q_{t}}{4\pi^2} \frac{v^2}{m_{d_i}}
    \sum_{k=1}^3 \frac{m_{t}}{M_{h_k}^2} \\
& \times \bigg\{
  \bigg[ Q_{t} f_{1}(x_{t h_k})
         - \frac{v_{u}^Z v_d^Z}{4Q_d} f_{3}(x_{t h_k}, x_{t Z}) \bigg]
  \text{Im}( \rho_{u,33}^* q_{k2} )
  \Big( \frac{m_{d_i}}{v} q_{k1} + \text{Re} (\rho_{d,ii}^* q_{k2}) \Big) \\
& \qquad - \bigg[ Q_{t} f_{2}(x_{t h_k})
                 - \frac{v_{u}^Z v_d^Z}{4Q_d} f_{4}(x_{t h_k}, x_{t Z}) \bigg]
  \text{Im}( \rho_{d,ii}^* q_{k2} )
  \Big( \frac{m_{t}}{v} q_{k1} + \text{Re} (\rho_{u,33}^* q_{k2}) \Big)
  \bigg\} \,,
\end{split}
\end{equation}
Here, $v_f^Z \equiv (T_3^f - 2Q_f s_w^2)/s_wc_w$ is the vectorial $Z$
coupling to fermion $f$, and $T_3^f = \pm 1/2$ is the weak isospin of
the fermion $f$. The result for the up-quark EDM, $C_3^{u,t h}$, can
be obtained by the replacements $m_{d_i} \to m_u$, $Q_{d} \to Q_u$,
$v_{d}^Z \to v_{u}^Z$, and $\rho_{d,ii} \to \rho_{u,11}$, and changing
the overall sign of the last line. The dimensionless two-loop
functions appearing in this result are well-known~\cite{Brod:2022bww};
they are given explicitly by
\begin{align}
  f_{1}(x) & = \Phi\bigg(\frac{1}{4x}\bigg) \,,
& f_{2}(x) & = (1-2x) f_{1}(x) + 2 \big(2 + \log(x) \big) \,, \label{eq:f12} \\
  f_{3}(x,y) & = \frac{y}{x - y} \bigg[ f_{1}(x) - f_{1}(y) \bigg] \,,
& f_{4}(x,y) & = \frac{y}{x - y}
                  \bigg[ f_{2}(x) - f_{2}(y) \bigg] \,. \label{eq:f34}
\end{align}
The function $\Phi(x)$ is defined in Ref.~\cite{Davydychev:1992mt}.

All charged SM fermions other than the top quark will also contribute
to the up- and down-quark EDMs. Inspection of the loop functions shows
that they are, for small fermion masses, numerically dominated by
powers of large logarithms of the form $\log(m_f^2/M_{h_k}^2)$; these
logarithms need to be summed to all orders in the strong coupling
constant. By contrast, naively using~\eqref{eq:C3di} in
the limit of small fermion masses would lead to wrong results. This
is explained in detail in Sec.~\ref{sec:light}.

\bigskip

Next, we discuss the contribution of diagrams with internal fermions
and charged Higgs bosons (see Fig.~\ref{fig:BZ:fermion}, left
panel). The result for the down-quark dipole is given, for internal
top and bottom quarks:
\begin{equation}\label{eq:dEDM:tH}
\begin{split}
  C_3^{d_i, tH^\pm}
= \frac{\alpha\alpha_s}{16\pi^2 s_w^2 Q_d} \frac{m_t v^2}{m_{d_i} M_{H^\pm}^2}
  \bigg( & \text{Im} \big( \rho_{u,33} \rho_{d,ii}^* \big) f_{5}(x_{t H^\pm}, x_{t W}) \\
& \quad + \frac{m_b}{m_t} \text{Im} \big( \rho_{d,33} \rho_{d,ii}^* \big)
          f_{6}(x_{t H^\pm}, x_{t W})\bigg) \,.
\end{split}
\end{equation}
Here, the explicit expressions for the two-loop functions are
\begin{equation}\label{eq:f5}
  f_{5}(x,y) = \frac{y}{x - y} \bigg[ \tilde f_{5}(x) - \tilde f_{5}(y) \bigg] \,,
\end{equation}
where
\begin{equation}
  \tilde f_{5}(x) = 3 x - \frac{1}{2}(13-6x)\log x
                  + (2-x)(2-3x)\bigg[ \text{Li}_2(1-1/x) - \frac{\pi^2}{6} \bigg] \,,
\end{equation}
and
\begin{equation}\label{eq:f6}
  f_{6}(x,y) = \frac{y}{x - y} \bigg[ \tilde f_{6}(x) - \tilde f_{6}(y) \bigg] \,,
\end{equation}
with
\begin{equation}
  \tilde f_{6}(x) = -3 x + \frac{1}{2}(1-6x)\log x
                  +x(2-3x)\bigg[ \text{Li}_2(1-1/x) - \frac{\pi^2}{6} \bigg] \,.
\end{equation}
The contribution to the down-quark EDM of diagrams with internal
charged leptons is calculated by expanding to linear order in the
lepton mass before performing the loop integrals. We find
\begin{equation}\label{eq:C3d:l}
  C_3^{d_i,\ell_j H^\pm} = \frac{\alpha\alpha_s}{32\pi^2 s_w^2 Q_d}
             \frac{m_{\ell_j} v^2}{m_{d_i} M_{H\pm}^2}
             \text{Im} \big( \rho_{\ell,jj} \rho_{d,ii}^* \big)
             \frac{1}{1 - x_{W H^\pm}} \log x_{W H^\pm} \,.
\end{equation}

The contribution of diagrams with a top or charm quark and charged
Higgs bosons to the up-quark EDM is given by
\begin{align}
  C_3^{u, tH^\pm}
& = - \frac{\alpha\alpha_s}{16\pi^2 s_w^2 Q_u} \frac{m_t v^2}{m_u M_{H^\pm}^2}
    \text{Im} \big( \rho_{u,33} \rho_{u,11}^* \big) f_{6}(x_{t H^\pm}, x_{t W}) \,,
    \label{eq:uEDM:tH} \\
  C_3^{u, cH^\pm}
& = - \frac{\alpha\alpha_s}{32\pi^2 s_w^2 Q_u} \frac{m_c v^2}{m_u M_{H^\pm}^2}
    \text{Im} \big( \rho_{u,22} \rho_{u,11}^* \big) \frac{\log x_{W H^\pm}}{x_{W H^\pm} - 1} \,.
    \label{eq:uEDM:cH}
\end{align}
In the diagrams with charged Higgs bosons, the virtual charm-quark
contributions to the down-quark EDM $C_3^{d_i}$, as well as the
virtual lepton and bottom-quark contributions to the up-quark EDM
$C_3^u$, lead to a large logarithm in the loop function that needs to
be summed to all orders, similar to the case of diagrams with neutral
Higgs bosons and light fermions, as discussed above. This is explained in
detail in the following section.

\subsection{Logarithmically enhanced light-fermion contributions to quark EDMs}\label{sec:light}

The two-loop top-quark contribution~\eqref{eq:C3di} to the quark EDMs
can, in principle, also be evaluated for the other quarks and charged
leptons. However, in the limit of small fermion masses, the loop
functions in Eqs.~\eqref{eq:f12} and~\eqref{eq:f34} contain large
logarithms that needs to be properly summed to all orders in the
strong coupling constant. One complication in this summation is
that, for the photon contribution in Eq.~\eqref{eq:C3di}, this
logarithm arises from QED operator mixing. For the $Z$ contribution,
on the other hand, the logarithm arises from one-loop initial
conditions to the tensor operators~\eqref{eq:op:qq':3:4}
and~\eqref{eq:op:semileptonic}, and subsequent QCD mixing into the
dipole operator. Similarly, a $W$ box contribution to the tensor
operators and subsequent QCD mixing into the dipole operator gives
rise to a large logarithm in the Barr-Zee diagrams with charged Higgs
bosons and virtual bottom quarks and leptons (for $C_3^{u}$), and with
virtual charm quarks (for $C_3^{d_i}$). In the following, we discuss
all these cases.

\subsubsection{Photon contribution}

\begin{figure}[t]
  \centering
  \includegraphics[width=12em]{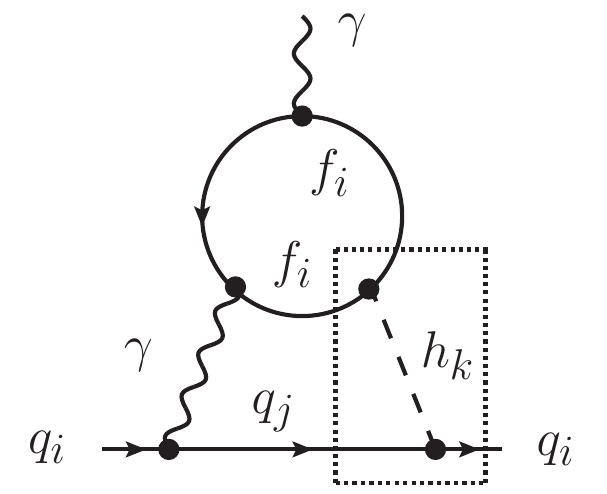}~~~~~~~~
  \includegraphics[width=9em]{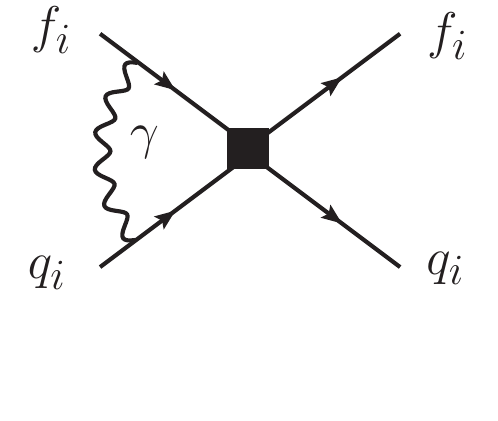}~~~~~~~~
  \includegraphics[width=9em]{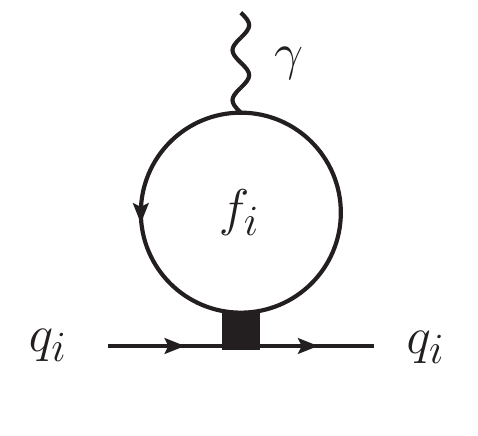}
  \caption{Integrating out the neutral Higgs bosons at tree level
    generates scalar four-fermion operators. They mix into tensor
    operators via one-loop photon exchange, and the tensor operators
    mix into the electric dipole operators. This generates the leading
    quadratic logarithms in Eq.~\eqref{eq:edm:leptons}. QCD RG
    evolution then sums the leading logarithms to all orders in
    $\alpha_s$. \label{fig:eft:photon}}
\end{figure}

The two-loop fermionic contributions to the quark EDMs, evaluated for
internal lepton loops, would take the form
\begin{equation}
\begin{split}\label{eq:edm:leptons}
  & C_3^{d_i,\ell_j h}
  = - \frac{\alpha\alpha_s}{4\pi^2} \frac{v^2}{m_{d_i}}
    \sum_{k=1}^3 \frac{m_{\ell_j}}{M_{h_k}^2} \\
& \times \bigg\{
  \bigg[ f_{1}(x_{\ell h_k})
         + \frac{v_{\ell}^Z v_d^Z}{4Q_d} f_{3}(x_{\ell_j h_k}, x_{\ell_j Z}) \bigg]
  \text{Im}( \rho_{\ell,jj}^* q_{k2} )
  \Big( \frac{m_{d_i}}{v} q_{k1} + \text{Re} (\rho_{d,ii}^* q_{k2}) \Big) \\
& \qquad + \bigg[ f_{2}(x_{\ell h_k})
                 + \frac{v_{\ell}^Z v_d^Z}{4Q_d} f_{4}(x_{\ell_j h_k}, x_{\ell_j Z}) \bigg]
  \text{Im}( \rho_{d,ii}^* q_{k2} )
  \Big( \frac{m_{\ell_j}}{v} q_{k1} + \text{Re} (\rho_{\ell,jj}^* q_{k2}) \Big)
  \bigg\} \,.
\end{split}
\end{equation}
(As before, the results for the up-quark EDM, $C_3^{u,\ell_j h}$, can
be obtained by the replacements $m_{d_i} \to m_u$, $Q_{d} \to Q_u$,
$v_{d}^Z \to v_{u}^Z$, and $\rho_{d,ii} \to \rho_{u,11}$, and changing
the overall sign of the last line.) The leading asymptotic values of
the loop functions for $x \ll 1$ and $y \ll 1$ are given
by\footnote{The loop functions for small fermion masses are
  numerically dominated by the leading logarithms. The next-leading
  terms in the expansion are $\pi^2/3$ for $f_{1}(x)$, a relative
  correction of 0.5\%, 1.6\%, 4.5\% for an internal electron, muon,
  and tau, respectively; and $2\log(x)$ for $f_{2}(x)$, corresponding
  to a relative correction of the order of ten percent. (Here we
  assume that $M_{h_k} \geq 125\,$GeV for all three neutral Higgs
  bosons.) The subleading corrections to $f_{3}(x,y)$ and $f_{4}(x,y)$
  are tiny. In all cases, it is a good approximation to keep only the
  (RG-improved) leading-logarithmic contribution.}
\begin{align} 
  f_{1}(x) & \simeq \log^2 x \,,
& f_{2}(x) & \simeq \log^2 x \,, \label{eq:limit:f12} \\
  f_{3}(x,y) & \simeq \frac{y}{x - y} \big[ \log x + \log y \big] \log \frac{x}{y} \,,
& f_{4}(x,y) & \simeq \frac{y}{x - y} \big[ \log x + \log y \big] \log \frac{x}{y} \,.
  \label{eq:limit:f34}
\end{align}
The large (linear and quadratic) logarithms should be summed to all
orders in the strong coupling $\alpha_s$. The photon and $Z$
contributions are treated quite differently in the effective theory
below the weak scale. The photon exchange generates quadratic
logarithms $\log^2(m_{\ell_j}^2/M_{h_k}^2)$, via the mixing of the
scalar operators $Q_1^{\ell_j q_i}$ and $Q_1^{q_i \ell_j}$ into the
tensor operators $Q_2^{\ell_j q_i}$ through single-photon exchange,
and subsequent mixing of $Q_2^{\ell_j q_i}$ into the electric dipole
$Q_e^{q_i}$. This mixing chain is illustrated in
Fig.~\ref{fig:eft:photon}. The linear large logarithms
$\log(m_{\ell_j}^2/M_{h_k}^2)$ and $\log(m_{\ell_j}^2/M_Z^2)$ of the
$Z$ contribution, on the other hand, arise from a one-loop initial
condition of the tensor operators $Q_2^{\ell_j q_i}$ at the weak
scale, and subsequent mixing of $Q_2^{\ell_j q_i}$ into the electric
dipole $Q_e^{q_i}$, see Fig.~\ref{fig:eft:WZ}.

We start by discussing the photon case. After the RG evolution, the
large logarithms will have the form $\log(\mu_\text{had}^2/M_{h_k})$,
with scale $\mu_\text{had}$ of the order of $2\,$GeV where the
hadronic matrix elements are evaluated. One sees immediately that a
naive evaluation of the loop functions in the fixed-order
result~\eqref{eq:edm:leptons} can lead to a gross overestimation of
the contribution: the result is ``wrong'' by a factor
$\log^2[(2\,\text{GeV})^2/m_\mu^2] \approx 35$ for the muon and
$\log^2[(2\,\text{GeV})^2/m_e^2] \approx 275$ for the
electron.\footnote{The electron case is phenomenologically not very
relevant, as the electron-Higgs couplings receive much stronger
constraints from bounds on the electron EDM, discussed in
Ref.~\cite{Altmannshofer:2024mbj}.} Moreover, this neglects the large
QCD effects that arise from the summation of the large logarithms.

The proper way of summing the large logarithms proceeds along the
lines presented in Ref.~\cite{Brod:2023wsh}. The rationale of this
calculation is the following. The limiting values of $f_1$ and $f_2$
in Eq.~\eqref{eq:limit:f12} can be reproduced as the leading term in a
series that has the schematic form
$\alpha \alpha_s \log^2(x_{\ell_j h_k}) (1 + \alpha_s \log(x_{\ell_j
  h_k}) + \alpha_s^2 \log^2(x_{\ell_j h_k}) + \ldots)$. Here, the
first term reflects the quadratic logarithm in
Eq.~\eqref{eq:limit:f12}, while all other terms correspond to the
leading logarithms of the diagrams obtained by dressing the Barr-Zee
diagrams, Fig.~\ref{fig:BZ:fermion}, with an arbitrary number of
virtual gluons. We use the formalism of Ref.~\cite{Huber:2005ig} to
resum the large logarithms $\alpha_s \log(x_{\ell_j h_k})$ to all
orders.

The required entries of the QED anomalous dimensions are given by
(cf. Ref.~\cite{Brod:2023wsh})
\begin{equation}
  \gamma_{C_1^{\ell q} \to C_2^{\ell q}}
= \gamma_{C_1^{q \ell} \to C_2^{\ell q}}
= \frac{\alpha}{2\pi} Q_q + \ldots \,.
\end{equation}
Other entries of the semileptonic anomalous dimension matrix are
either zero or numerically irrelevant. The ellipsis denotes QCD and
higher-order contributions. The exact leading-logarithmic result for
charged-lepton contributions to the electric dipole Wilson coefficient
is simple enough to be written out in the following compact form:
\begin{equation}
\begin{split}
  C_3^{q_i, \ell_j h}(\mu_\text{had})
                & = \bigg[ \frac{12}{115} \eta_3^{-11/27} \eta_4^{-9/25} \eta_5^{-7/23}
                    - \frac{144}{4025} \eta_3^{-11/27} \eta_4^{-9/25} \eta_5^{16/23} \\
                & \qquad 
                    - \frac{16}{525} \eta_3^{-11/27} \eta_4^{16/25} \eta_5^{16/23}
                    - \frac{8}{21} \eta_3^{16/27} \eta_4^{16/25} \eta_5^{16/23} \\
                & \qquad 
                    + \frac{48}{2975} \eta_3^{-11/27} \eta_4^{-9/25} \eta_5
                    + \frac{32}{4725} \eta_3^{-11/27} \eta_4^{16/25} \eta_5 \\
                & \qquad 
                    + \frac{16}{189} \eta_3^{16/27} \eta_4^{16/25} \eta_5
                    + \frac{80}{8721} \eta_3^{-11/27} \eta_4 \eta_5 \\
                & \qquad 
                    + \frac{16}{297} \eta_3^{16/27} \eta_4 \eta_5
                    + \frac{36}{209} \eta_3 \eta_4 \eta_5 \bigg]
                \frac{\alpha}{\alpha_s^{(5)}(m_t)}
                \frac{m_{\ell_j}}{v}
                \big[ C_1^{\ell_j q_i}(m_t) + C_1^{q_i \ell_j}(m_t) \big] \,,
\end{split}
\end{equation}
where $\eta_5 = \alpha_s^{(5)}(m_t)/\alpha_s^{(5)}(m_b)$,
$\eta_4 = \alpha_s^{(4)}(m_b)/\alpha_s^{(4)}(m_c)$, and
$\eta_3 = \alpha_s^{(3)}(m_c)/\alpha_s^{(3)}(\mu_\text{had})$; the
superscript denotes the number of active quark flavors. The initial
conditions for the Wilson coefficients are given in
Eqs.~\eqref{eq:C1eq} and~\eqref{eq:C1qe}, below. It is straightforward to
check that this solution exactly reproduces the quadratic large
logarithm in Eq.~\eqref{eq:edm:leptons}.

The bottom and charm contributions to the light-quark EDMs are
obtained in a completely analoguous way. The only required, non-zero
entries of the QED anomalous dimension matrix are
\begin{equation}
  \gamma_{C_1^{q q'} \to C_3^{q q'}}
= \gamma_{C_1^{q' q} \to C_3^{q q'}}
= - \frac{\alpha}{2\pi} Q_q Q_{q'} + \ldots \,.
\end{equation}
Again, all other entries are not needed, and we indicate QCD and
higher-order terms by the ellipsis. The initial conditions are
required for the coefficients $C_1^{ub}$, $C_1^{bu}$, $C_1^{d_i d_3}$,
$C_1^{d_3 d_i}$ for virtual bottom quarks, and $C_1^{uc}$, $C_1^{cu}$,
$C_1^{d_i c}$, $C_1^{c d_i}$ for virtual charm quarks. They are given
explicitly in Eqs.~\eqref{eq:C1ud},~\eqref{eq:C1du},
and~\eqref{eq:C1qiqj}.\footnote{Here and in the following, we do not
need to consider the contributions of virtual up, down, and strange
quarks, as their couplings receive much stronger contraints from
contributions to the up-, down-, and strange-quark EDMs arising from
{\em heavy} virtual particles (e.g. via top-loop or bosonic Barr-Zee
diagrams).}

The structure of the anomalous dimensions for virtual bottom and charm
quarks is more complicated than in the semi-leptonic case, and it is
not instructive to present the analytic result explicitly. Needless to
say, the corresponding solution of the RG equation is included in the
python code.

\boldmath
\subsubsection{$Z$- and $W$-boson contributions}
\unboldmath

\begin{figure}[t]
  \centering
  \includegraphics[width=12em]{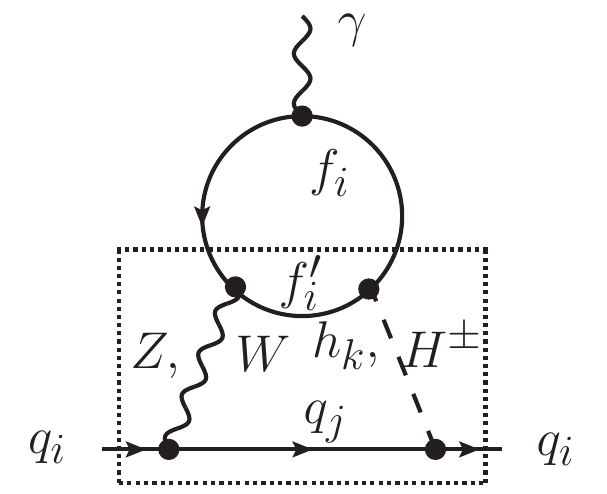}~~~~~~~~
  \includegraphics[width=9em]{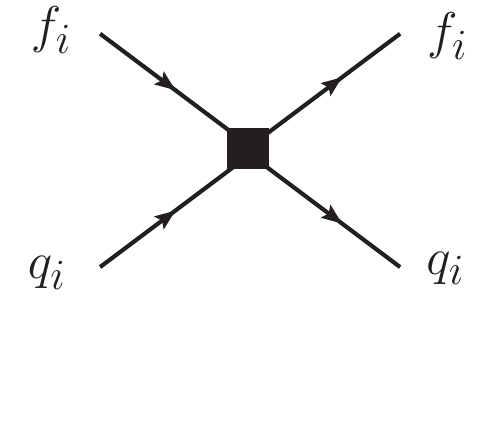}~~~~~~~~
  \includegraphics[width=9em]{./figs/edmA}
  \caption{Integrating out the Higgs, $W$, and $Z$ bosons at one-loop
    generates tensor four-fermion operators. They mix into the
    electric dipole operators. This generates the leading linear
    logarithms in Eq.~\eqref{eq:edm:leptons}. QCD RG evolution then
    sums the leading logarithms to all orders in
    $\alpha_s$. \label{fig:eft:WZ}}
\end{figure}

The leading-logarithmic contribution of virtual $Z$ bosons to the
quark EDMs with virtual bottom, charm, and lepton loops is obtained
via the one-loop initial conditions for the four-quark and
semileptonic tensor operators, given in Eqs.~\eqref{eq:C3ud},
\eqref{eq:C3ub}, \eqref{eq:C3qq}, \eqref{eq:C2lu},
and~\eqref{eq:C2ld}, respectively, and subsequent QCD RG evolution
(mixing of the tensor operators into the electric dipole
operators). The ``mixing chain'' is explained in
Fig.~\ref{fig:eft:WZ}.

Similarly, one obtains the leading-logarithmic contribution of virtual
$W$ bosons to the quark EDMs with virtual bottom, charm, and lepton
loops via the one-loop initial conditions for the four-quark and
semileptonic tensor operators, given in Eqs.~\eqref{eq:C3ud},
\eqref{eq:C3ub}, \eqref{eq:C2lu}, and~\eqref{eq:C2ld},
respectively. The exceptions are the bottom and lepton contributions
to the down- and strange-quark EDMs, and the charm contribution to the
up-quark EDM. For these particular flavor combinations, the tensor
operators do not mix into the dipole operators, and no large logarithm
appears in the two-loop functions. The two-loop results can then
directly be used as initial conditions for the RG evolution; the
corresponding expressions have been given in
Eqs.~\eqref{eq:dEDM:tH},~\eqref{eq:uEDM:cH}, and~\eqref{eq:C3d:l}.

\subsubsection{One-loop contributions}

The function $g_1$ appearing in the one-loop top-quark
contributions~\eqref{eq:C34u} and~\eqref{eq:C3d} to the quark EDMs is
also proportional to a large logarithm for the case of internal bottom
and charm quarks, corresponding to the left and right panels in
Fig.~\ref{fig:edm:1loop} with charm and bottom quarks instead of top
quark. (Also the corresponding diagrams with external down and strange
quarks need to be taken into account.) This logarithm is reproduced
and summed to all orders by the RG evolution, starting with the
tree-level initial conditions~\eqref{eq:C3ud},~\eqref{eq:C3ub},
and~\eqref{eq:C3qq} for the tensor four-quark operators, and
subsequent mixing into the electric dipole operators.

\subsection{Bosonic contributions to the quark EDMs}\label{sec:bosonic}

In this section we present the contributions from diagrams involving
vertices arising from the kinetic terms of the Higgs bosons and the
Higgs potential. Among the diagrams with vertices from the Higgs
kinetic terms only, there are two classes that are separately gauge
invariant. The first class consists of the diagrams with internal $Z$
and neutral Higgs bosons that are not of the Barr-Zee type (see
Fig.~\ref{fig:BZ:fermion}, right panel):
\begin{equation}\label{eq:edm:hZ}
\begin{split}
  C_3^{d_i, hZ} & = \quad \frac{e\alpha\alpha_s \big(v_d^Z\big)^2}{96\pi^2 s_w c_w} \frac{v^2}{m_{d_i} M_Z}
             \sum_{k=2}^3 q_{k1} \text{Im} \big( \rho_{d,ii}^* q_{k2} \big)
             \big[ f_{7+}(x_{Z h_k}) - f_{7+}(x_{Z h_1}) \big] \\
          & \quad + \frac{e\alpha\alpha_s \big(a_d^Z\big)^2}{96\pi^2 s_w c_w} \frac{v^2}{m_{d_i} M_Z}
             \sum_{k=2}^3 q_{k1} \text{Im} \big( \rho_{d,ii}^* q_{k2} \big)
             \big[ f_{7-}(x_{Z h_k}) - f_{7-}(x_{Z h_1}) \big] \,,
\end{split}
\end{equation}
where $a_f^Z \equiv - T_3^f/(s_wc_w)$ is the axial $Z$ coupling to
fermion $f$. The loop functions are
\begin{equation} \label{eq:f27pm}
\begin{split}
  f_{7\pm}(x) & = \frac{16 x^4 + 4 x^3 - (2\pm 24)x^2 \pm (18 x - 3)}{4x^3} \Phi\bigg(\frac{1}{4x}\bigg)
                 + \frac{8 x^2 - 2 x \mp 3}{2x} \\
              & \quad - \frac{4 x^5 + 3 x^4 - (1\pm 3) x^2 \pm (12 x - 3)}{x^3} \, \text{Li}_2 (1 - x)
                      - \frac{8 x^5 + 6 x^4 \mp (12 x - 3)}{12 x^3} \pi^2 \\
              & \quad - \log^2(x) \frac{8 x^5 + 6 x^4 - (2\pm 6)x^2 \pm (12 x - 3)}{4x^3}
                      + \log(x) \frac{4 x^2 + x \pm 3}{x} \,.
\end{split}
\end{equation}
The dilogarithm is defined as
\begin{equation}
  \text{Li}_2 (x) = -\int_0^x du \, \log (1-u)/u \,.
\end{equation}
The result for the up-quark EDM, $C_3^{u, hZ}$, can be obtained by the
replacements $m_{d_i} \to m_u$, $v_{d}^Z \to v_{u}^Z$,
$a_{d}^Z \to a_{u}^Z$, and $\rho_{d,ii} \to \rho_{u,11}$, and changing
the overall sign.

\begin{figure}[t]
  \centering
  \includegraphics[width=12em]{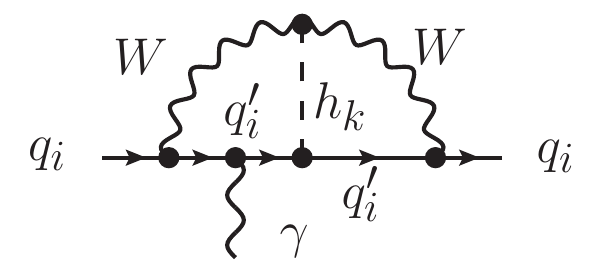}~~~
  \includegraphics[width=12em]{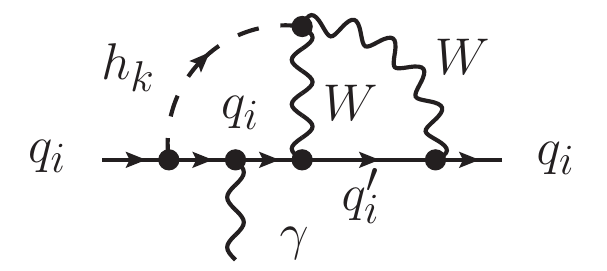}~~~
  \includegraphics[width=12em]{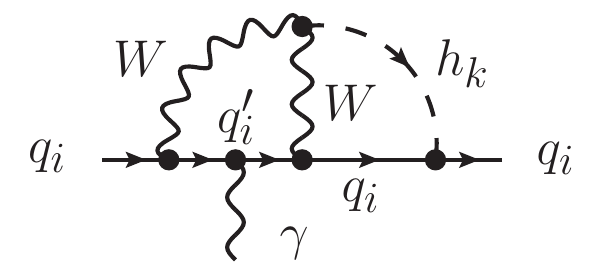}~~~
  \caption{Representative non-Barr-Zee two-loop diagrams with $W$ exchange
    contributing to the light-quark EDM. The external photon can be
    attached to any internal charged line.  The four diagrams represented
    by the left panel have no correspondence in the electron EDM case;
    they turn out to be zero.
    \label{fig:higgs:kinetic}}
\end{figure}

All remaining diagrams with vertices from the Higgs kinetic terms give
\begin{equation}\label{eq:eEDM:bosonic:d}
\begin{split}
  C_3^{d_i,\text{kin.}} & = \quad \frac{e\alpha\alpha_s v_d^Z}{64\pi^2 Q_d} \frac{v^2}{m_{d_i}M_Z}
             \sum_{k=2}^3 q_{k1} \text{Im} \big( \rho_{d,ii}^* q_{k2} \big)
             \big[ f_{12a}(x_{Z h_k}, c_w^2) - f_{12a}(x_{Z h_1}, c_w^2) \big] \\
                     & \quad - \frac{e\alpha\alpha_s}{32 \pi^2 s_w} \frac{v^2}{m_{d_i}M_W}
             \sum_{k=2}^3 q_{k1} \text{Im} \big( \rho_{d,ii}^* q_{k2} \big)
             \big[ f_{8c}(x_{W h_k}) - f_{8c}(x_{W h_1}) \big] \\
                     & \quad + \frac{e\alpha\alpha_s}{144 \pi^2 s_w^3 Q_d} \frac{v^2}{m_{d_i}M_W}
             \sum_{k=2}^3 q_{k1} \text{Im} \big( \rho_{d,ii}^* q_{k2} \big)
             \big[ f_{12b}(x_{W h_k}) - f_{12b}(x_{W h_1}) \big] \\
                     & \quad + \frac{e\alpha\alpha_s}{64 \pi^2 Q_d s_w^3} \frac{v^2}{m_{d_i}M_W}
             \sum_{k=2}^3 q_{k1} \text{Im} \big( \rho_{d,ii}^* q_{k2} \big)
             \big[ f_{8e}(x_{W h_k}, x_{W H^+}) - f_{8e}(x_{W h_1}, x_{W H^+}) \big] \,.
\end{split}
\end{equation}
The four lines in Eq.~\eqref{eq:eEDM:bosonic:d} correspond to
Barr-Zee-type diagrams (one Higgs and one gauge boson coupling to the
external quark) with internal $Z$-boson exchange (first line),
Barr-Zee-type diagrams with internal photon exchange (second line),
diagrams with internal $W$-boson exchange that are not of the Barr-Zee
type (one Higgs and two gauge bosons coupling to the external quark,
see diagrams in Fig.~\ref{fig:higgs:kinetic} -- third line),
Barr-Zee-type diagrams with internal $W$-boson exchange (fourth
line). Note that this splitting is not entirely physical, as it
depends on the choice of gauge parameter. Only in the sum of all
contributions and after using the orthogonality conditions of the
Higgs mixing angles the gauge parameters drop out, as we verified
explicitly (see also App.~\ref{sec:gauge}). The gauge-independent
parts of the loop functions are given by (the somewhat idiosyncratic
naming of the loop functions is to avoid inconsistency with the naming
of loop functions in Ref.~\cite{Altmannshofer:2024mbj})
\begin{align}
\begin{split} \label{eq:f12a}
  f_{12a}(x,y) & =   \frac{12 x y^2 - 16 x y + 3 x}{(1-x)(1-y)} \, \Phi\bigg(\frac{1}{4y}\bigg)
                  - \frac{12 x^2 y^2 - 16 x y + 3}{(1-x)(1-y)} \Phi\bigg(\frac{1}{4xy}\bigg) \\
& \quad           + \log(x) \frac{12 x y^2 - 2 x y + 2 y - 1}{(1-x)(1-y)y} \,,
\end{split}
\\
\begin{split} \label{eq:f8c}
f_{8c}(x) & = \big(12 x^2 - 16 x + 3 \big) \Phi\bigg(\frac{1}{4x}\bigg)
               - 2 \big[ \log(x) + 2 \big] \big(6 x + 1\big)
\end{split}
\\
\begin{split} \label{eq:f12b}
  f_{12b}(x) & = \frac{4 x^2 + 3 x}{12} \pi^2
          + \log^2(x) \frac{4 x^3 + 3 x^2 - 1}{x}
          - \log(x) \big( 2x + 5 \big) \\
& \quad 
          + \frac{8 x^4 + 6 x^3 - 2 x - 9}{4 x^2} \text{Li}_2(1 - x)\\
& \quad 
          + \frac{38 x^3 - 49 x^2 - 16 x + 9}{8 x^2} \Phi\bigg(\frac{1}{4x}\bigg)
          - \frac{8 x - 11}{4} \,,
\end{split}
\\
\begin{split} \label{eq:f8e}
f_{8e}(x,y) & = \log^2(x) \frac{x^3 y (3 - y)  
          + 3 x^2 y^2 - x (3y^2 + 4y +1) + y^2 + y}{2 x^3}\\
& \quad + \log(x) \log(y) \frac{x^3 y (y^2 - 4 y + 3) - 3 x^2 y^2 (y - 1)
                                + x y^2 (3y+1) - y^3}{2 x^3 (y - 1)}\\
& \quad 
+ \log(x) \frac{y - x y - x}{x^2}
+ \log(y) \frac{x^2 (1 + 7 y - 2 y^2) + xy (4y+1) - 2 y^2}{2 x^2 (y - 1)}\\
& \quad
+ \bigg[ \frac{ x^4 y (3 - y^3 + 5 y^2 - 7 y) + x^3 y^3 ( 4 y - 8)}{2 x^4 (y - 1)}\\
& \quad \qquad
- \frac{x^2 y^2 (6 y^2 - y + 1) - x y^3 (4 y + 2) + y^4}{2 x^4 (y - 1)}
  \bigg] \Phi\bigg(y, \frac{y}{x}\bigg)\\
& \quad 
+ \frac{ x^3 y (3 - y) + 3 x^2 y^2 - x y (3 y + 4) + y (y + 1)}{x^3} \text{Li}_2(1 - x)\\
& \quad 
+ \frac{4 x^3 + 6 x^2 - 6 x + 1}{2 x^3 (y - 1)} \, y \, \Phi\bigg(\frac{1}{4x}\bigg)
+ \frac{x y (2 - x) + x - y}{x^2} \,.
\end{split}
\end{align}
The function $\Phi(x,y)$ is defined in
Ref.~\cite{Davydychev:1992mt}. For the up-quark EDM we find
\begin{equation}\label{eq:eEDM:bosonic:u}
\begin{split}
  C_3^{u,\text{kin.}} & = - \frac{e\alpha\alpha_s v_u^Z}{64\pi^2 Q_u} \frac{v^2}{m_{u}M_Z}
             \sum_{k=2}^3 q_{k1} \text{Im} \big( \rho_{u,ii}^* q_{k2} \big)
             \big[ f_{12a}(x_{Z h_k}, c_w^2) - f_{12a}(x_{Z h_1}, c_w^2) \big] \\
                   & \quad + \frac{e\alpha\alpha_s}{32 \pi^2 s_w} \frac{v^2}{m_{u}M_W}
            \sum_{k=2}^3 q_{k1} \text{Im} \big( \rho_{u,ii}^* q_{k2} \big)
            \big[ f_{8c}(x_{W h_k}) - f_{8c}(x_{W h_1}) \big] \\
                   & \quad + \frac{e\alpha\alpha_s}{144 \pi^2 s_w^3 Q_u} \frac{v^2}{m_{u} M_W}
             \sum_{k=2}^3 q_{k1} \text{Im} \big( \rho_{u,ii}^* q_{k2} \big)
             \big[ f_{12c}(x_{W h_k}) - f_{12c}(x_{W h_1}) \big] \\
                   & \quad + \frac{e\alpha\alpha_s}{64 \pi^2 Q_u s_w^3} \frac{v^2}{m_{u}M_W}
             \sum_{k=2}^3 q_{k1} \text{Im} \big( \rho_{u,ii}^* q_{k2} \big)
             \big[ f_{8e}(x_{W h_k}, x_{W H^+}) - f_{8e}(x_{W h_1}, x_{W H^+}) \big] \,,
\end{split}
\end{equation}
where
\begin{equation}
\begin{split} \label{eq:12c}
  f_{12c}(x) & = \frac{4 x^2 + 3 x}{6} \pi^2
          + \log^2(x) \frac{4 x^3 + 3 x^2 - 1}{2x}
          - \log(x) \frac{8x + 11}{2} \\
& \quad 
          + \frac{16 x^4 + 12 x^3 - 4 x - 9}{4 x^2} \text{Li}_2(1 - x)\\
& \quad 
          + \frac{22 x^3 - 53 x^2 - 14 x + 9}{8 x^2} \Phi\bigg(\frac{1}{4x}\bigg)
          - \frac{16 x - 13}{4} \,.
\end{split}
\end{equation}

\begin{figure}[t]
  \centering
  \includegraphics[width=12em]{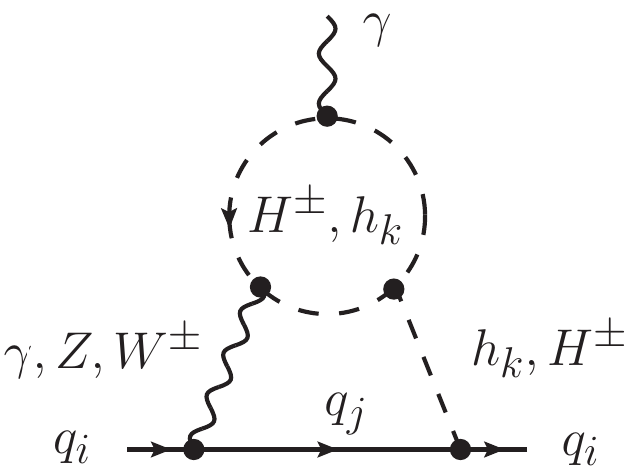}~~~
  \includegraphics[width=12em]{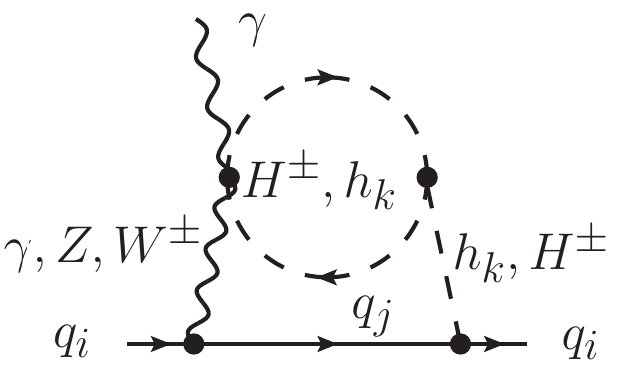}~~~
  \includegraphics[width=12em]{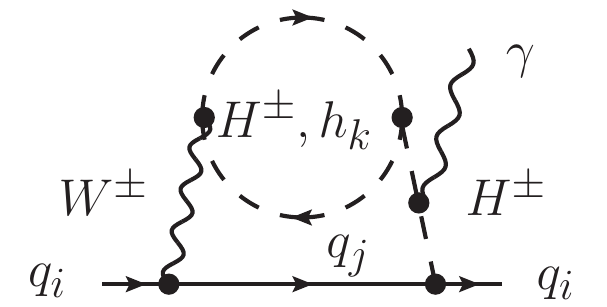}
  \caption{Representative two-loop diagrams contributing to the
    light-quark EDM with vertices arising from the Higgs potential. 
    \label{fig:higgs:potential}}
\end{figure}

Last, we discuss the contributions of diagrams with vertices derived
from the Higgs potential. Such contribution necessarily enter at the
two-loop level (see Fig.~\ref{fig:higgs:potential}). The contribution
of diagrams with internal photons and charged Higgs bosons is
\begin{equation}
 C_3^{d_i, H^\pm \gamma} = - \frac{\alpha\alpha_s}{8\pi^2}
             \sum_{k=1}^3 \frac{v^3}{m_{d_i} M_{h_k}^2} \text{Im} \big( \rho_{d,ii}^* q_{k2} \big)
                  \Big[ q_{k1} Z_3 + \text{Re}(q_{k2} Z_7)
                  \Big] f_{9}(x_{H^\pm h_k}) \,,
\end{equation}
with the loop function
\begin{equation} \label{eq:f29}
  f_{9}(x) = x \, \Phi\bigg(\frac{1}{4x}\bigg) - \log(x) - 2 \,.
\end{equation}
The contribution of diagrams with internal $Z$ bosons and charged
Higgs bosons is
\begin{equation}
 C_3^{d_i,H^\pm Z} = \frac{\alpha\alpha_s}{32\pi^2} \frac{c_w^2-s_w^2}{s_w c_w Q_d} v_d^Z
             \sum_{k=1}^3 \frac{v^3}{m_{d_i} M_{h_k}^2} \text{Im} \big( \rho_{d,ii}^* q_{k2} \big)
                  \Big[ q_{k1} Z_3 + \text{Re}(q_{k2} Z_7)
                  \Big] f_{10}(x_{H^\pm h_k}, x_{Z h_k}) \,,
\end{equation}
with
\begin{equation} \label{eq:f210}
  f_{10}(x,y) = \frac{1}{1 - y} \bigg[ \log(y) - x \, \Phi\bigg(\frac{1}{4x}\bigg)
                                       + \frac{x}{y} \Phi\bigg(\frac{y}{4x}\bigg) \bigg] \,.
\end{equation}
The contribution of diagrams with internal $W$ bosons and charged
Higgs bosons is
\begin{equation}
 C_3^{d_i,H^\pm W} = \frac{\alpha\alpha_s}{64\pi^2 s_w^2 Q_d}
             \sum_{k=1}^3 \frac{v^3}{m_{d_i} M_{h_k}^2} \text{Im} \big( \rho_{d,ii}^* q_{k2} \big)
                  \Big[ q_{k1} Z_3 + \text{Re}(q_{k2} Z_7)
                  \Big] f_{11}(x_{H^\pm h_k}, x_{W h_k}) \,,
\end{equation}
with
\begin{equation} \label{eq:f211}
\begin{split}
  f_{11}(x,y) & = \log(x) \log(y) \frac{y x - x^2 + 2 x - 1}{y^2(y-x)}
+ \log(y) \frac{2 x - y - 2}{y(y-x)} \\
& \quad + \log^2(x) \frac{2 x - 1}{x^2(x-y)}
+ \log(x) \frac{y x + 2 y - 2 x^2}{xy(y-x)}
+ \frac{2(x - 1)}{y x} \\
& \quad 
+ \frac{2 y x^2 - y^2 x  - y x - y - 
         x^3 + 3 x^2 - 3 x + 1}{y^2x(y-x)} \Phi\bigg(\frac{y}{x}, \frac{1}{x}\bigg) \\
& \quad 
+ \frac{2(y - 2 y x + x^3 - 2 x^2 + x)}{y^2 x^2} \text{Li}_2(1 - x)
+ \frac{2 x^2 - 4 x + 1}{x^2(x-y)} \Phi\bigg(\frac{1}{4x}\bigg) \,.
\end{split}
\end{equation}
The three contributions are separately gauge invariant. The
contribution of all diagrams with internal $Z$ bosons and vertices
from the Higgs potential that do not contain charged Higgs bosons are
zero. The results for the up-quark EDM, $C_3^{u, H^\pm \gamma}$,
$C_3^{u,H^\pm Z}$, and $C_3^{u,H^\pm W}$, can be obtained by the
replacements $m_{d_i} \to m_u$, $Q_{d_i} \to Q_u$, $v_{d}^Z \to
v_{u}^Z$, and $\rho_{d,ii} \to \rho_{u,11}$, and changing the overall
sign for $C_3^{u, H^\pm \gamma}$ and $C_3^{u,H^\pm Z}$, {\em but not
  for} $C_3^{u,H^\pm W}$.

\subsection{Contributions to the quark chromo-EDMs}\label{sec:cedm}

\begin{figure}[t]
  \centering
  \includegraphics[width=12em]{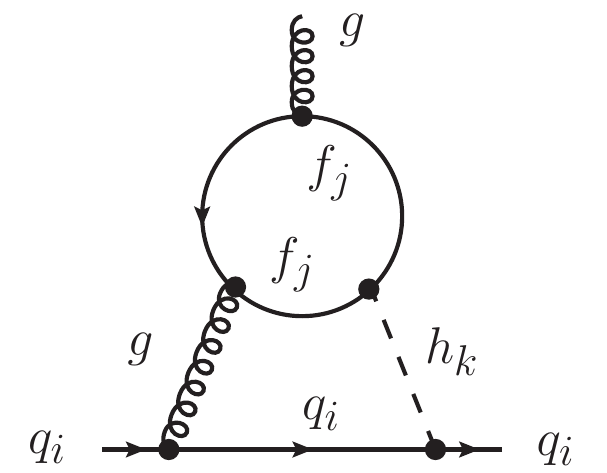}~~~
  \includegraphics[width=12em]{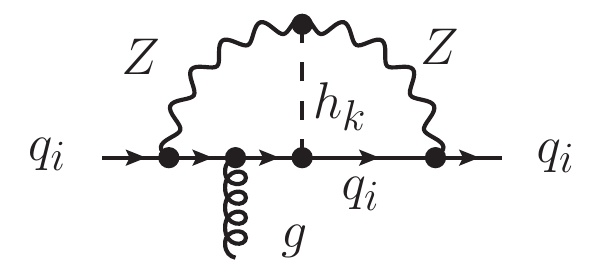}
  \caption{Representative two-loop diagrams contributing to the
    light-quark chromo-EDM. \label{fig:CEDM:BZ:fermion}}
\end{figure}

Here, we present the initial conditions for the chromo-EDMs of the up,
down, strange, charm and bottom quarks. The charm and bottom
chromo-EDMs are needed because they contribute to the coefficient of
the Weinberg operator via finite threshold corrections, see
Eq.~\eqref{eq:th:wbo}. It is useful to split $C_4^{q_i}$ into a
fermionic and a bosonic contribution,
$C_4^{q_i} = C_4^{q_i,\text{fermionic}} +
C_4^{q_i,\text{bosonic}}$. We find, for $d_1 = d$, $d_2 = s$,
$d_3 = b$,
\begin{equation}
\begin{split} \label{eq:C4d:fermion}
  C_4^{d_i,\text{fermionic}} & = \frac{\alpha_s}{2\pi} \frac{m_t}{m_{d_i}} \frac{v^2}{M_{H^\pm}^2}
               g_1(x_{tH^\pm}) \text{Im} \big( \rho^*_{u,i3} \rho_{d,i3} \big) \\
  & \quad - \frac{\alpha_s^2}{8\pi^2} \frac{m_{t}}{m_{d_i}}
    \sum_{k=1}^3 \frac{v^2}{M_{h_k}^2}
 \bigg\{ \quad
           f_{1}(x_{t h_k})
  \text{Im}( \rho_{u,33}^* q_{k2} )
  \Big( \frac{m_{d_i}}{v} q_{k1} + \text{Re} (\rho_{d,ii}^* q_{k2}) \Big) \\
& \hspace{10.5em} - f_{2}(x_{t h_k})
  \text{Im}( \rho_{d,ii}^* q_{k2} )
  \Big( \frac{m_{t}}{v} q_{k1} + \text{Re} (\rho_{u,33}^* q_{k2}) \Big)
  \bigg\} \,.
\end{split}
\end{equation}
For the up-quark ($u_1 = u$, $u_2 = c$) chromo-electric dipoles, we
find
\begin{equation}
\begin{split} \label{eq:C4u:fermion}
  C_4^{u_i,\text{fermionic}} & = - \frac{\alpha_s}{4\pi} \frac{m_t}{m_{u_i}}
               \sum_k \frac{v^2}{M_{h_k}^2} g_1(x_{t h_k})
               \text{Im} \big\{\rho_{u,3i}^* \rho_{u,i3}^* q_{k2}^2 \big\} \\
  & \quad - \frac{\alpha_s^2}{8\pi^2} \frac{m_{t}}{m_{u_i}}
    \sum_{k=1}^3 \frac{v^2}{M_{h_k}^2}
 \bigg\{ \quad
           f_{1}(x_{t h_k})
  \text{Im}( \rho_{u,33}^* q_{k2} )
  \Big( \frac{m_{u_i}}{v} q_{k1} + \text{Re} (\rho_{u,ii}^* q_{k2}) \Big) \\
& \hspace{10.5em} + f_{2}(x_{t h_k})
  \text{Im}( \rho_{u,ii}^* q_{k2} )
  \Big( \frac{m_{t}}{v} q_{k1} + \text{Re} (\rho_{u,33}^* q_{k2}) \Big)
  \bigg\} \,.
\end{split}
\end{equation}
The first lines in Eqs.~\eqref{eq:C4d:fermion}
and~\eqref{eq:C4u:fermion} correspond to the one-loop contribution
(see middle panel in Fig.~\ref{fig:edm:1loop}); the second and third
lines represent the contributions of gluonic Barr-Zee diagrams with
internal top-quark loops and neutral Higgs bosons (see
Fig.~\ref{fig:CEDM:BZ:fermion}, left panel). They can be obtained by
simple replacements of couplings and charges (and dropping the $Z$
contribution) from the results for the electric dipole,
Eq.~\eqref{eq:C3di}. The dimensionless two-loop functions appearing in
this result have been given above, Eq.~\eqref{eq:f12}.

It should be obvious that these contributions to the chromo-EDMs of
the light quarks are numerically much larger than the corresponding
contributions (with photons instead of gluons) to the electric dipole
moments. In fact, the contributions to the electric dipole moments
that are induced, via QCD RG running, from chromo-electric dipole
moments typically dominate by large factors over the direct
contributions, discussed in the previous sections. Note, however, that
diagrams with virtual leptons or with vertices from the Higgs
potential do not contribute to the chromo-EDMs and are only tested via
the electric dipole moments.

\bigskip

The result for the non-Barr-Zee $Z$ and $W$ contributions is
\begin{equation}\label{eq:cedm:W}
  C_4^{d_i,\text{bosonic}} =  - C_3^{d_i, hZ}
                           + \frac{\alpha\alpha_s e}{96\pi^2s_w^3} \frac{v^2}{M_W m_{d_i}}
             \sum_{k=1}^3 q_{k1} \text{Im} \big( \rho_{d,ii}^* q_{k2} \big)
             f_{13}(x_{W h_k}) \,,
\end{equation}
where the loop function is
\begin{equation} \label{eq:f13}
\begin{split}
  f_{13}(x) & = 1 - 4 x - (4 x + 1) \log(x)
               - \frac{8 x^2 + 2 x - 1}{2x} \Phi\bigg(\frac{1}{4x}\bigg) \\
           & \quad + \frac{4 x^3 + 3 x^2 - 1}{2x} \, \big[\log^2(x) + 2 \text{Li}_2 (1 - x) \big]
               + \frac{4 x^2 + 3x}{6} \pi^2 \,.
\end{split}
\end{equation}
The result for the up- and charm-quark EDMs,
$C_4^{u_i,\text{bosonic}}$, can be obtained by the replacements
$C_3^{d_i, hZ} \to C_3^{u_i, hZ}$, and changing $m_{d_i} \to m_u$,
$\rho_{d,ii} \to \rho_{u,11}$, as well as the overall sign, in the
second term.

\subsubsection*{Light-quark contributions}

Contributions of virtual quarks other than the top are taken into
account via the mixing of scalar and tensor four-quark operators into
the chromo-electric dipole, in analogy to the case of electric
dipoles, described in detail in Sec.~\ref{sec:light}. In particular,
the large quadratic logarithm that appears in the two-loop
contribution in Eqs.~\eqref{eq:C4d:fermion} and~\eqref{eq:C4u:fermion}
when replacing the top quark with bottom or charm quarks can be
reproduced and summed to all orders in QCD, using the initial
conditions for the scalar four-quark operators,
Eqs.~\eqref{eq:C1ud},~\eqref{eq:C1du}, and~\eqref{eq:C1qiqj}, and
subsequent mixing into the tensor and chromo-electric dipole
operators. This chain can again be illustrated with
Fig.~\ref{fig:eft:photon}, where one should now replace all photons by
gluons.\footnote{Higher-order resummation of these logarithms has been
  performed in Ref.~\cite{Brod:2018pli}.}

Similarly, the large logarithm in the one-loop contribution in
Eqs.~\eqref{eq:C4d:fermion} and~\eqref{eq:C4u:fermion}, evaluated for
virtual bottom and charm quarks, is reproduced and summed to all
orders in QCD, using the initial conditions for the tensor four-quark
operators, starting with the tree-level initial
conditions~\eqref{eq:C3ud},~\eqref{eq:C3ub}, and~\eqref{eq:C3qq}, and
subsequent mixing into the chromo-electric dipole operators.

\subsection{Contributions to the Weinberg operator}\label{sec:wbo}

\begin{figure}[t]
  \centering
  \includegraphics[width=10em]{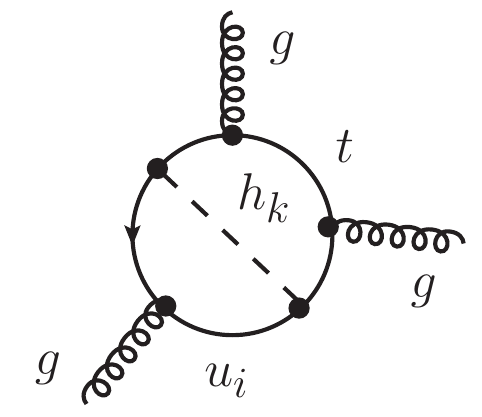}~~~~~~~
  \includegraphics[width=10em]{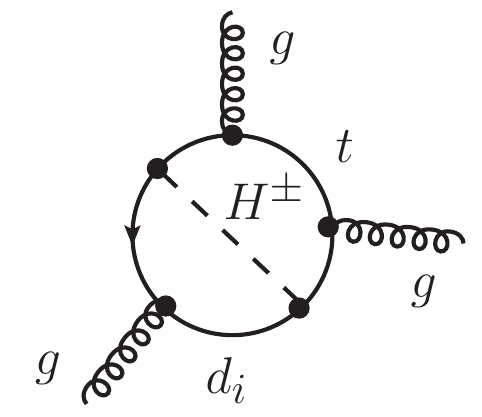}
  \caption{Representative two-loop diagrams contributing to the gluon
    chromo-EDM (``Weinberg operator''). \label{fig:WBO}}
\end{figure}

The contribution of the Weinberg operator to hadronic EDMs is
generally small~\cite{Engel:2013lsa}. Its significance arises from the
fact that it receives contributions that are proportional to the
(modified) top-quark Yukawa coupling only. These contributions arise
from two-loop diagrams involving virtual top quarks, see
Fig.~\ref{fig:WBO}. The corresponding initial conditions are
\begin{equation}\label{eq:wbo}
\begin{split}
  C_w^{h} & = - \frac{\alpha_s^2}{32\pi^2} \sum_{k} \frac{v^2}{M_{h_k}^2}
          \Big[   \text{Im} \big( (\rho_{u,33}^*)^2 q_{k2}^2 \big)
                + \frac{2m_t}{v} q_{k1} \text{Im} \big( \rho_{u,33}^* q_{k2} \big)
          \Big] f_{14}(x_{t h_k}) \\
      & \quad - \frac{\alpha_s^2}{32\pi^2} \sum_{k} \frac{v^2}{M_{h_k}^2} \sum_{i=1,2}
          \Big[   \text{Im} \big( \rho_{u,i3}^* \rho_{u,3i}^* q_{k2}^2 \big)
                + \frac{m_t}{v} q_{k1} \text{Im} \big( \rho_{u,i3}^* q_{k2} \big)
          \Big] f_{15}(x_{t h_k}) \\
      & \quad - \frac{\alpha_s^2}{16\pi^2} \frac{v^2}{M_{H^\pm}^2} \sum_{i=1}^3
                   \text{Im} \big( \rho_{d,i3}^* \rho_{u,i3} \big)
                   f_{15}(x_{t H^\pm})
              \,,
\end{split}
\end{equation}
where (cf. the results in Ref.~\cite{Brod:2022bww})
\begin{align}
  f_{14}(x) & = \frac{6x^2-5x-1}{(4x-1)^3} \log x
               - \frac{2x-3}{2(4x-1)^2}
               - \frac{6x^3-6x^2+3x}{(4x-1)^3} \Phi(1/4x) \,, \\
  f_{15}(x) & = \frac{x}{(x-1)^4} \log x
               + \frac{x^2-5x-2}{6(x-1)^3} \,.
\end{align}
Note that the Weinberg operator receives additional logarithmically
enhanced contributions from finite threshold corrections at the
bottom- and charm-quark scales. Again, the most significant are those
arising from the bottom-quark contribution to the bottom-quark
chromo-EDM, and the charm-quark contribution to the charm-quark
chromo-EDM, see Sec.~\ref{sec:rge}. In analogy to the top-quark
contributions, they are respectively proportional to modifications of
the bottom- and charm-quark Yukawas only. They are additional finite
contributions, at the weak scale as well as at the bottom and charm
thresholds, arising from two-loop diagrams with virtual light quarks,
analoguos to those shown in Fig.~\ref{fig:WBO} (and corresponding
diagrams where the Higgs propoagators are replaced by insertions of
effective four-quark operators). All these contributions are
proportional to combinations of couplings that are already probed by
contributions to the quark (chromo-)EDMs that have much larger
hadronic matrix elements. Hence, we neglect these contributions to the
Weinberg operator.

\subsection{Contributions to four-quark operators}\label{sec:four:quark}

We calculate the matching conditions at the weak scale for all
four-quark operators in our basis. While their direct contribution to
hadronic CP violation is negligible (their hadronic matrix elements
are tiny, see Ref.~\cite{Engel:2013lsa}), they mix into the
(chromo-)electric dipole operators and thus account for the CP
violation induced by virtual light quarks, as explained in
Sec.~\ref{sec:light}.

The following Wilson coefficients receive their leading contributions
at tree-level:
\begin{align}
  C_1^{u_i d_j} & = - \sum_k \frac{v^2}{M_{h_k}^2}
              \Big( \frac{m_{u_i}}{v} q_{k1} + \text{Re}\big(\rho_{u,ii}^* q_{k2}\big) \Big)
              \text{Im}\big(\rho_{d,jj}^* q_{k2}\big)
              - \frac{1}{6} \frac{v^2}{M_{H^\pm}^2}
              \text{Im} \big( \rho^*_{d,ji} \rho_{u,ji} \big) \,, \label{eq:C1ud} \\
  C_1^{d_i u_j} & = \sum_k \frac{v^2}{M_{h_k}^2}
              \Big( \frac{m_{d_i}}{v} q_{k1} + \text{Re}\big(\rho_{d,ii}^* q_{k2}\big) \Big)
              \text{Im}\big(\rho_{u,jj}^* q_{k2}\big)
              - \frac{1}{6} \frac{v^2}{M_{H^\pm}^2}
              \text{Im} \big( \rho^*_{d,ij} \rho_{u,ij} \big) \,, \label{eq:C1du} \\
  C_2^{u_i d_j} & = C_2^{d_j u_i} = 2 C_4^{u_i d_j} = - \frac{v^2}{M_{H^\pm}^2}
              \text{Im} \big( \rho^*_{d,ji} \rho_{u,ji} \big) \,.
\end{align}

The operator $Q_3^{u_i d_j}$, on the other hand, receives
contributions both at tree-level and at one-loop. The relevance of the
one-loop contributions has been explained in
Sec.~\ref{sec:light}. Note in particular that the tree-level and
one-loop contributions depend on different Higgs coupling
combinations. For $j=1,2$, we find
\begin{equation} \label{eq:C3ud}
\begin{split}
  C_3^{u_i d_j} & = - \frac{v^2}{12M_{H^\pm}^2}
              \text{Im} \big( \rho^*_{d,ji} \rho_{u,ji} \big) \\
              & \quad + \frac{\alpha}{16\pi s_w^2} \frac{v^2}{M_{H\pm}^2}
               \text{Im} \big( \rho_{u,ii} \rho_{d,jj}^* \big)
               \frac{\log x_{W H^\pm}}{x_{W H^\pm} - 1} \\
              & \quad + \frac{\alpha v_u^Z v_d^Z}{16\pi}
                     \sum_k \frac{v^2}{M_{h_k}^2}
                     \frac{\log x_{Z h_k}}{1 - x_{Z h_k}} \\
              & \qquad \times
                     \bigg[ \text{Im} \big( \rho_{u,ii} \rho_{d,jj}^* \big) | q_{k2}|^2
                            - \frac{m_{d_j}}{v} q_{k1} \text{Im} \big( \rho_{u,ii}^* q_{k2} \big)
                            + \frac{m_{u_i}}{v} q_{k1} \text{Im} \big( \rho_{d,jj}^* q_{k2} \big)
                     \bigg] \,,
\end{split}
\end{equation}
while for $d_j=b$, we find
\begin{equation} \label{eq:C3ub}
\begin{split}
  C_3^{u_i b} & = - \frac{v^2}{12M_{H^\pm}^2}
              \text{Im} \big( \rho^*_{d,3i} \rho_{u,3i} \big) \\
              & \quad + \frac{\alpha}{16\pi s_w^2} \frac{v^2}{M_{H\pm}^2}
               \text{Im} \big( \rho_{u,ii} \rho_{d,33}^* \big) f_{16}(x_{W H^\pm}, x_{t H^\pm}) \\
              & \quad + \frac{\alpha v_u^Z v_d^Z}{16\pi}
                     \sum_k \frac{v^2}{M_{h_k}^2}
                     \frac{\log x_{Z h_k}}{1 - x_{Z h_k}} \\
              & \qquad \times
                     \bigg[ \text{Im} \big( \rho_{u,ii} \rho_{d,33}^* \big) | q_{k2}|^2
                            - \frac{m_{b}}{v} q_{k1} \text{Im} \big( \rho_{u,ii}^* q_{k2} \big)
                            + \frac{m_{u_i}}{v} q_{k1} \text{Im} \big( \rho_{d,33}^* q_{k2} \big)
                     \bigg] \,,
\end{split}
\end{equation}
with the loop function
\begin{equation}
  f_{16}(x,y) = \frac{x \log x}{x^2 - x y - x + y} + \frac{y \log y}{y^2 - x y + x - y} \,.
\end{equation}
Furthermore, for $i \neq j$ we find
\begin{align}
  C_1^{q_i q_j} & = \pm \sum_k \frac{v^2}{M_{h_k}^2} \bigg\{
                  \Big( \frac{m_{q_i}}{v} q_{k1} + \text{Re}(\rho_{q,ii}^* q_{k2}) \Big)
                  \text{Im}(\rho_{q,jj}^* q_{k2})
                  + \frac{1}{12} \text{Im} \big( \rho_{q,ij}^* \rho_{q,ji}^* q_{k2}^2 \big)
                  \bigg\} \,, \label{eq:C1qiqj} \\
  C_2^{q_i q_j} & = C_2^{q_j q_i} = 2 C_4^{q_i q_j}
                = \mp \frac{1}{2} \sum_k \frac{v^2}{M_{h_k}^2}
                  \text{Im} \big( \rho_{q,ij}^* \rho_{q,ji}^* q_{k2}^2 \big) \,,
                  \label{eq:C2qiqj} \\
\begin{split} \label{eq:C3qq}
    C_3^{q_i q_j} & = \mp \frac{1}{24} \sum_k \frac{v^2}{M_{h_k}^2}
                  \text{Im} \big( \rho_{q,ij}^* \rho_{q,ji}^* q_{k2}^2 \big) \\
                & \quad \pm \frac{\alpha \big(v_q^Z\big)^2}{16\pi}
                     \sum_k \frac{v^2}{M_{h_k}^2}
                     \frac{\log x_{Z h_k}}{x_{Z h_k} - 1} \\
                 & \qquad \times
                     \bigg[ \text{Im} \big( \rho_{q,ii}^* \rho_{q,jj}^* \big) | q_{k2}|^2
                            + \frac{m_{q_j}}{v} q_{k1} \text{Im} \big( \rho_{q,ii}^* q_{k2} \big)
                            + \frac{m_{q_i}}{v} q_{k1} \text{Im} \big( \rho_{q,jj}^* q_{k2} \big)
                     \bigg] \,.
\end{split}
\end{align}
Here, the upper signs are for $q_i = u_i$ and the lower signs for
$q_i = d_i$. And finally, for the operators with four equal quarks, we
have
\begin{align}
  C_1^{q_i} & = \pm \sum_k \frac{v^2}{M_{h_k}^2}
                 \Big( \frac{m_{q_i}}{v} q_{k1} + \text{Re} (\rho_{q,ii}^* q_{k2}) \Big)
                 \text{Im}\big\{ \rho_{q,ii}^* q_{k2} \big\} \,, \label{eq:C1qi}
\end{align}
with, again, the upper sign for $q_i = u_i$ and the lower sign for
$q_i = d_i$. The coefficient $C_2^{q_i}$ receives a contribution from
box diagrams with virtual $Z$ and neutral Higgs bosons, but they are
not phenomenologically relevant.

To obtain some of the results in this section, we used the Fierz
identities collected in App.~\ref{sec:fierz}. The naive application of
these four-dimensional relations is permitted, since we are only
calculating leading-logarithmic effects.

\subsection{Contributions to semi-leptonic operators}\label{sec:semi:leptonic}

Semi-leptonic four-fermion operators play a two-fold role in our
analysis. They directly induce CP-violating electron-quark
interactions, which translate into CP-violating electron-nucleon
interactions that can be probed by measuring the EDMs of paramagnetic
atoms and molecules~\cite{Engel:2013lsa}. In addition, they mix into
the quark electric dipole operators and thus account for the CP
violation induced by virtual charged leptons, as explained in
Sec.~\ref{sec:light}.

The matching calculation at the weak scale gives the following
results:
\begin{align}
  C_1^{\ell_i q_j} & = \pm \sum_k
              \Big( \frac{v^2}{M_{h_k}^2} \frac{m_{\ell_i}}{m_{q_j}} q_{k1}
                    + \frac{v^3}{M_{h_k}^2 m_{q_j}} \text{Re}(\rho_{\ell,ii}^* q_{k2}) \Big)
              \text{Im}(\rho_{q,jj}^* q_{k2}) \,, \label{eq:C1eq} \\
  C_1^{q_i \ell_j} & = - \sum_k
              \Big( \frac{v^2}{M_{h_k}^2} q_{k1}
                    + \frac{v^3}{M_{h_k}^2 m_{q_i}} \text{Re}(\rho_{q,ii}^* q_{k2}) \Big)
              \text{Im}(\rho_{\ell,jj}^* q_{k2}) \,. \label{eq:C1qe}
\end{align}
Here, the upper sign applies for $q_i = u_i$ and the lower sign for
$q_i = d_i$. At one-loop, also the tensor operators get generated:
\begin{align}
\begin{split} \label{eq:C2lu}
  C_2^{\ell_i u_j} & = \frac{\alpha}{16\pi s_w^2} \frac{v^3}{M_{H\pm}^2 m_{u_j}}
                    \text{Im} \big( \rho_{\ell,ii} \rho_{u,jj}^* \big)
                    \frac{\log x_{W H^\pm}}{1 - x_{W H^\pm}} \\
                & \quad + \frac{\alpha v_\ell^Z v_u^Z}{16\pi}
                     \sum_k \frac{v^3}{M_{h_k}^2 m_{u_j}}
                     \frac{\log x_{Z h_k}}{1 - x_{Z h_k}} \\
                 & \quad \times
                     \bigg[ \text{Im} \big( \rho_{\ell,ii}^* \rho_{u,jj} \big) | q_{k2}|^2
                            - \frac{m_{\ell_i}}{v} q_{k1} \text{Im} \big( \rho_{u,jj}^* q_{k2} \big)
                            + \frac{m_{u_j}}{v} q_{k1} \text{Im} \big( \rho_{\ell,ii}^* q_{k2} \big)
                     \bigg] \,,
\end{split}\\
\begin{split} \label{eq:C2ld}
  C_2^{\ell_i d_j} & = \frac{\alpha v_\ell^Z v_d^Z}{16\pi}
                     \sum_k \frac{v^3}{M_{h_k}^2 m_{d_j}}
                     \frac{\log x_{Z h_k}}{1 - x_{Z h_k}} \\
                 & \quad \times
                     \bigg[ \text{Im} \big( \rho_{\ell,ii}^* \rho_{d,jj}^* q_{k2}^2 \big)
                            + \frac{m_{\ell_i}}{v} q_{k1} \text{Im} \big( \rho_{d,jj}^* q_{k2} \big)
                            + \frac{m_{d_j}}{v} q_{k1} \text{Im} \big( \rho_{\ell,ii}^* q_{k2} \big)
                     \bigg] \,.
\end{split}
\end{align}
The initial condition for the CP-violating electron-gluon
operator~\eqref{eq:op:eeGG} arises at one-loop and is given by
\begin{equation}
  C_5^{e} = - \sum_k \left(   \frac{v^2}{M_{h_k}^2} q_{k1}
                           + \frac{v^3}{M_{h_k}^2 m_t} \text{Re}(\rho_{u,33}^* q_{k2}) \right)
             \text{Im} (\rho_{\ell,11}^* q_{k2}) \,.
\end{equation}

\section{Conclusion} \label{sec:conclusions}

We presented the leading contributions to the light-quark ($q=u,d,s$)
EDMs and chromo-EDMs, as well as to the Weinberg operator and CP-odd
semileptonic four-fermion operators, in the unconstrained 2HDM. We
kept terms that are at most quadratic in the Yukawa interactions. To
the best of our knowledge, these results have been presented here in
full generality for the first time. We paid particular attention to
consistently summing all large logarithms using QCD and QED RG
evolution using an low-energy effective theory. The corresponding RG
evolution of the Wilson coefficients is valid not only for the 2HDM,
but generally for any model that does not introduce new light degrees
of freedom beyond the SM.

We verified our results through several consistency checks. We
confirmed that all infrared and ultraviolat divergences cancel in our
results. Moreover, we performed the calculation in generalized $R_\xi$
gauge for all gauge bosons, and verified explicitly that all our
results are gauge parameter independent.

Our results can be used to calculate the contributions in the 2HDM to
EDMs of various nuclear, atomic, and molecular systems, see
Ref.~\cite{Engel:2013lsa}.

\subsection*{Python code}

To make it easier to use our results in numerical applications, we
provide an implementation of our results into \texttt{python} code. It
can be downloaded via the public git repository
\begin{center}
\url{https://gitlab.com/jbrod/general-2hdm-pheno}
\end{center}
and provides (in addition to the results of
Ref.~\cite{Altmannshofer:2024mbj}) the numerical values for all Wilson
coefficients at the hadronic scale $\mu=2\,$GeV, and the values of the
quark and gluon (chromo-)EDMs as defined in Sec.~\ref{sec:dipole}, for
user-specified model input parameters. Information on the usage of the
code can be found in the repository.

\section*{Acknowledgments}

J.B. acknowledges support by DoE grant DE-SC0011784. The research of W.A. is supported by the U.S. Department of Energy grant number DE-SC0010107. The work of P.U. is supported in part by Thailand NSRF via PMU-B under grant number B39G680009.

\appendix

\section{Fierz relations}\label{sec:fierz}

We used the following Fierz identities to project onto our basis:
\begin{align}
\begin{split}
       (\bar q q') \, (\bar q' \, i \gamma_5 q)
     + (\bar q' q) \, (\bar q \, i \gamma_5 q')
 & = - \frac{1}{6} \Big( Q_1^{qq'} + Q_1^{q'q} \Big)
     - \Big( Q_2^{qq'} + Q_2^{q'q} \Big) \\
 & \quad 
     - \frac{1}{12} Q_3^{qq'} - \frac{1}{2} Q_4^{qq'}
     - \frac{1}{12} Q_3^{qq'} - \frac{1}{2} Q_4^{qq'} \,,
\end{split}\\
\begin{split}
       (\bar q \, T^a  q') \, (\bar q' \, i \gamma_5 T^a q)
     + (\bar q' \, T^a  q) \, (\bar q \, i \gamma_5 T^a q')
 & = - \frac{2}{9} \Big( Q_1^{qq'} + Q_1^{q'q} \Big)
     + \frac{1}{6} \Big( Q_2^{qq'} + Q_2^{q'q} \Big) \\
 & \quad 
     - \frac{1}{9} Q_3^{qq'} + \frac{1}{12} Q_4^{qq'} \,,
\end{split}\\
\begin{split}
       \frac{1}{2}\epsilon^{\mu\nu\rho\sigma} (\bar q \sigma_{\mu\nu} q')
       \, (\bar q' \sigma_{\rho\sigma} q)
 & = - \Big(Q_1^{qq'} + Q_1^{q'q} \Big) - 6 \Big( Q_2^{qq'} + Q_2^{q'q} \Big) \\
 & \quad
     + \frac{1}{6} Q_3^{qq'} + Q_4^{qq'} \,,
\end{split}\\
\begin{split}
       \frac{1}{2}\epsilon^{\mu\nu\rho\sigma} (\bar q \sigma_{\mu\nu} T^a q')
       \, (\bar q' \sigma_{\rho\sigma} T^a q)
 & = - \frac{4}{3} \Big( Q_1^{qq'} + Q_1^{q'q} \Big)
     + \Big( Q_2^{qq'} + Q_2^{q'q} \Big) \\
 & \quad 
     + \frac{2}{9} Q_3^{qq'} - \frac{1}{6} Q_4^{qq'} \,.
\end{split}
\end{align}
The structures
$(\bar q q') \, (\bar q' \, i \gamma_5 q) - (\bar q' q) \, (\bar q \,
i \gamma_5 q')$ and
$(\bar q \, T^a q') \, (\bar q' \, i \gamma_5 T^a q) - (\bar q' \, T^a
q) \, (\bar q \, i \gamma_5 T^a q')$ are CP even and do not mix into
the CP-odd operators. Hence, they do not contribute to EDMs. For the
case of four equal quarks, we have two more relations:
\begin{eqnarray}
 (\bar q T^a q) \, (\bar q \, i \gamma_5 T^a q)
 &=& -\frac{5}{12} Q_1^{q} - \frac{1}{16} Q_2^{q} \,, \\
 \frac{1}{2}\epsilon^{\mu\nu\rho\sigma} (\bar q \sigma_{\mu\nu} T^a q) \, (\bar q \sigma_{\rho\sigma} T^a q)
 &=& -3 Q_1^{q} + \frac{1}{12} Q_2^{q} \,.
\end{eqnarray}

\section{Gauge-parameter dependent parts of the bosonic loop functions}\label{sec:gauge}

As the bosonic contributions to the quark EDMs have been calculated
here for the first time (part of the contributions can be extracted
from the results in Ref.~\cite{Altmannshofer:2020shb}), we show the
gauge-parameter dependent parts of the loop
functions~\eqref{eq:f12a}-\eqref{eq:f8e} explicitly. It is
straightforward to verify that all these contributions cancel in the
sum, after applying unitarity relations.
\begin{align}
\begin{split} \label{eq:f12axi}
  f_{12a,\xi}(x,y) & = \log(\xi) \log(x) \frac{ x y \xi (2 - \xi) - x y + \xi + 3}{y - 1}
                     + \log(\xi) \log(y) \frac{ x y \xi (2 - \xi) - x y}{y - 1} \\
& \quad
          + \frac{x^2 y^2 (\xi^3 - 3 \xi^2 + 3 \xi - 1)
                  - x y (2 \xi^2 + 2 \xi - 4) + \xi - 3}{y - 1}
            \Phi(x y,x y \xi) \\
& \quad 
          + \frac{2 x y \xi^2 - \xi}{y - 1} \Phi(1/4xy\xi)
          + \frac{2 x y (\xi^2 - 2 \xi + 1)}{y - 1} \text{Li}_2(1 - \xi) \,,
\end{split}
\\
\begin{split} \label{eq:f8cxi}
f_{8c,\xi}(x) & = \big(2 x \xi^2 - \xi \big) \Phi\bigg(\frac{1}{4x\xi}\bigg)
                 + 2 x \big(\xi^2 - 2 \xi + 1 \big) \text{Li}_2 (1-\xi) \\
& \quad + \big( x^2 \xi^3 - 3 x^2 \xi^2
          + 3 x^2 \xi - x^2 - 2 x\xi^2 - 2x\xi + 4x
          + \xi - 3 \big) \Phi\bigg(x,x\xi\bigg)  \\
& \quad - 2 \log\xi + \xi \log^2\xi
        - \big( x\xi^2 - 2x\xi + x - \xi - 3 \big) \log\xi \log x \,,
\end{split}
\\
\begin{split} \label{eq:f12bxi}
   f_{12b,\xi}(x) & = \quad 9 \log x \log \xi \frac{x^3\xi^2 - 2x^3\xi + x^3
          + x^2\xi^2 - 2x^2\xi - 3x^2 - 2x\xi + 1}{8x^2}\\
& \quad 
          + 9 \log^2(x) \frac{x\xi^2 - x\xi - 2\xi}{8 x}
          + \frac{9\xi}{8} \log^2\xi - (1+\log\xi) \frac{9\xi}{4} \\
& \quad 
          + 9 \frac{- x\xi^2 + 2x\xi - x}{4} \text{Li}_2(1 - \xi)
          + 9 \frac{x^2\xi^2 - x^2\xi - 2x\xi + 1}{4x^2} \text{Li}_2(1 - x\xi)\\
& \quad 
          + 9\frac{ - 2x\xi^2 + \xi}{8} \Phi\bigg(\frac{1}{4x\xi}\bigg)\\
& \quad 
          + 9 \bigg[ \frac{- 3x^2\xi^2 + 3x^2 + 3x\xi + x - 1}{8x^2} \\
& \quad \qquad + \frac{ - x^2\xi^3 + 3x^2\xi^2
          - 3x^2\xi + x^2 + x\xi^3 + 3x\xi - 4x}{8}
          \bigg] \Phi\bigg(x,x\xi\bigg) \,,
\end{split}
\\
\begin{split} \label{eq:f8exi}
f_{8e,\xi}(x,y) & = \log(x)^2 \frac{ - x \xi^2 + x \xi + 2 \xi}{2 x}
          + \log(\xi) \log(x) \frac{ - x^2 \xi^2 + x^2 \xi + 2 x \xi - 1}{2 x^2}\\
& \quad
          - \xi \log(\xi) - \xi
          + \text{Li}_2(1 - x \xi) \frac{ - x^2 \xi^2 + x^2 \xi + 2 
         x \xi - 1}{ x^2}\\
& \quad 
          + \Phi(x,x \xi) \frac{ - x^3 \xi^3 + 2 x^3 \xi^2
          - x^3 \xi + 3 x^2 \xi^2 - x^2 \xi - 3 x \xi - 
         x + 1}{2 x^2} \,,
\end{split}\\
   f_{12c,\xi}(x) & = f_{12b,\xi}(x) \,.
\end{align}

\addcontentsline{toc}{section}{References}
\bibliographystyle{JHEP}
\bibliography{references}{}

\end{document}